\documentclass[pdflatex,sn-nature]{gmmag}

\usepackage{graphicx}
\makeatletter
\let\GMMagSICaption\caption
\let\GMMagSIInternalCaption\@caption
\let\GMMagSIXFloat\@xfloat
\let\GMMagSIFloatReset\@floatboxreset
\makeatother
\usepackage{float}
\usepackage{amsmath,amssymb,amsfonts}
\usepackage{booktabs}
\usepackage{array}
\usepackage{longtable}
\usepackage{placeins}
\usepackage{xcolor}
\usepackage{textcomp}
\usepackage{bibunits}

\makeatletter
\newskip\GMMagFloatTop
\newskip\GMMagFloatBottom
\GMMagFloatTop=\@fptop
\GMMagFloatBottom=\@fpbot
\makeatother
\makeatletter
\def\au@and{\ifnum\punctcount=2\ and\else\unskip, \global\advance\punctcount by -1 \fi}
\makeatother

\hypersetup{filecolor=black}

\hypersetup{hypertexnames=false}
\newcommand{\MainEq}[1]{(\ref{#1})}
\DeclareRobustCommand{\suppref}[1]{\begingroup\hypersetup{linkcolor=black}\ref{#1}\endgroup}

\begin{document}

\makeatletter
\let\GMMagTitle\title
\let\GMMagAuthor\author
\let\GMMagMakeTitle\maketitle
\let\GMMagInternalMakeTitle\@maketitle
\makeatother

\begin{bibunit}[sn-nature]

\title[GMMag]{Magnetic crystal generation with site-resolved magnetic moments via Gaussian-mixture retargeting and finite-budget yield prediction}

\author[1]{\fnm{Wen-Chuang} \sur{Shang}}
\author[1]{\fnm{Chenyu} \sur{Shen}}
\author[1]{\fnm{Mikito} \sur{Tsujimoto}}
\author[1]{\fnm{Tomonori} \sur{Tanaka}}
\author*[1]{\fnm{Yoshihiro} \sur{Gohda}}

\affil[1]{\orgdiv{Department of Materials Science and Engineering}, \orgname{Institute of Science Tokyo}, \orgaddress{\city{Yokohama} \postcode{226-8501}, \country{Japan}}\\
\textcolor{blue}{\href{mailto:shang.w.beed@m.isct.ac.jp}{shang.w.beed@m.isct.ac.jp}}; \textcolor{blue}{\textsuperscript{*}\href{mailto:gohda@mct.isct.ac.jp}{gohda@mct.isct.ac.jp}}}

\abstract{We introduce GMMag, a workflow for generating crystal structures together with their site-resolved magnetic moments, built on a conditional variational autoencoder with a graph neural network encoder and a transformer decoder. Gaussian-mixture sampling redirects generation towards selected chemical and magnetic targets without retraining. We systematically investigate how encoder and decoder design, the representation of magnetic-moment magnitudes and relative signs, and sampling choices affect reconstruction, generation quality and diversity. Because a higher screening pass rate does not necessarily yield more qualifying candidates with distinct chemical formulas absent from the reference dataset, we model how repeated generation of the same chemical formulas limits their accumulation and test the model in two separate ten-million-draw runs. As an application, the workflow identifies Nd-containing candidates with chemical formulas absent from the reference dataset, negative formation energies and high estimated saturation magnetization.}


\maketitle

\section{Introduction}

Magnetic materials underpin permanent magnets, spintronic devices and energy-conversion technologies. Their inverse design requires control over composition, crystal structure and local spin configuration. High-throughput repositories and graph neural networks have expanded the structures that can be evaluated computationally \cite{jain2013materialsproject,kirklin2015oqmd,curtarolo2012aflow,xie2018cgcnn,choudhary2021alignn}, while generative models provide distributions from which new candidates can be sampled. For magnetic crystals, a small net magnetic moment can describe either weak local magnetism or large antiparallel local moments that cancel. A net-moment label alone does not distinguish these cases. Representing site-resolved magnetic moments retains their relative orientations; sums of atom-projected moments must also be distinguished from the magnetic moment of the complete cell.

Variational, autoregressive and diffusion models, including CDVAE \cite{xie2022cdvae}, CrystaLLM \cite{antunes2024crystallm}, CrystalFormer \cite{cao2025crystalformer}, MatterGen \cite{zeni2025mattergen} and Chemeleon \cite{park2025chemeleon}, now generate crystals with increasingly diverse compositions and properties. Magnetic applications have combined such models with screening and first-principles calculations for magnetocaloric and spintronic candidates \cite{court2021magnetocaloric,siriwardane2023spintronic}, permanent magnets using MagGen \cite{mal2024maggen}, antiferromagnetic candidates \cite{gong2025antiferromagnets}, and rare-earth-free magnets using DiffCrysGen \cite{mal2026diffcrysgen}. Complementary first-principles studies of Nd magnets have addressed candidate discovery through database screening and transfer-learning-assisted genetic searches \cite{seo2024ndfe16,ito2026ndfe}, as well as lattice dynamics and their dependence on magnetic order \cite{tsuna2023phonon,gohda2021intergranular}. Target control can enter through conditional training, as in FTCP \cite{ren2022ftcp}, Cond-CDVAE \cite{luo2024condcdvae} and Con-CDVAE \cite{ye2024concdvae}, reward-directed training in Crystal-GFN \cite{hernandezgarcia2023crystalgfn}, or predictor-guided sampling in PODGen \cite{ye2026podgen}. These routes differ in their reliance on learned conditions, evaluator calls and downstream screening.

Fitting a sampling distribution after training offers a complementary route when suitable reference structures are available in the training data. Gaussian mixture models (GMMs) support controlled molecular sampling in CLaSS \cite{das2021class}, latent sampling in Dis-GEN \cite{petersen2025disgen} and target-subset exploration for organic semiconductors \cite{nakanishi2026organicgmm}, as well as output distributions in CrystalFormer \cite{cao2025crystalformer}, feature encoding \cite{ridwan2026legoxtal} and density-constrained optimization \cite{bajpai2026vibann}. A trained encoder maps selected reference structures to compact numerical representations, or latent vectors. Replacing the distribution fitted to these vectors can redirect generation without changing the generator weights or optimizing each generated candidate. The relevant question is whether this approach can jointly target chemical composition and magnetic properties while retaining local magnetic-moment information.

Evaluating this capability also requires separating immediate success from accumulated discovery. A high screening pass rate can repeatedly recover the same chemical formulas. Even a formula absent from the fixed reference set contributes only once to a generation run's distinct discoveries. Sampling choices must therefore be judged by both the probability of passing the selection criteria and the number of distinct qualifying reference-unseen formulas found. Sparse site-resolved magnetic labels create a further constraint: predicted labels can extend training coverage, but agreement with generated moments remains evidence from a predictive model.

Here we introduce GMMag, a conditional variational autoencoder that represents signed local magnetic moments alongside crystal structure. Its training set combines structures from public repositories with a smaller set from in-house density functional theory (DFT) calculations. When DFT-derived site-resolved magnetic moments are available, either from public records or the in-house calculations, they are used directly; for public structures without such labels, a structure-based predictor trained beforehand on labelled crystals supplies estimated site-resolved moments. The trained generator is then redirected towards selected targets through GMM sampling without retraining. We evaluate generation tasks defined by chemical-composition constraints and magnetic-moment criteria. We then model how distinct qualifying reference-unseen formulas accumulate with the number of generation attempts, testing the predictions against independently generated long runs. Models trained from scratch under matched conditions separately test decoder capacity, conditioning paths, additional magnetic supervision and the input paths carrying magnetic-moment signs. Finally, relaxation with a machine-learning interatomic potential (MLIP), DFT and atomic site-occupancy calculations evaluate selected Nd-containing candidates. These experiments assess model design, magnetic representation, adaptation to different targets and discovery efficiency.

\section{Results}

\subsection{Site-resolved magnetic moments in crystal generation}\label{sec:site-moment-results}

\begin{figure}[tbp]
\centering
\includegraphics[width=0.99\textwidth]{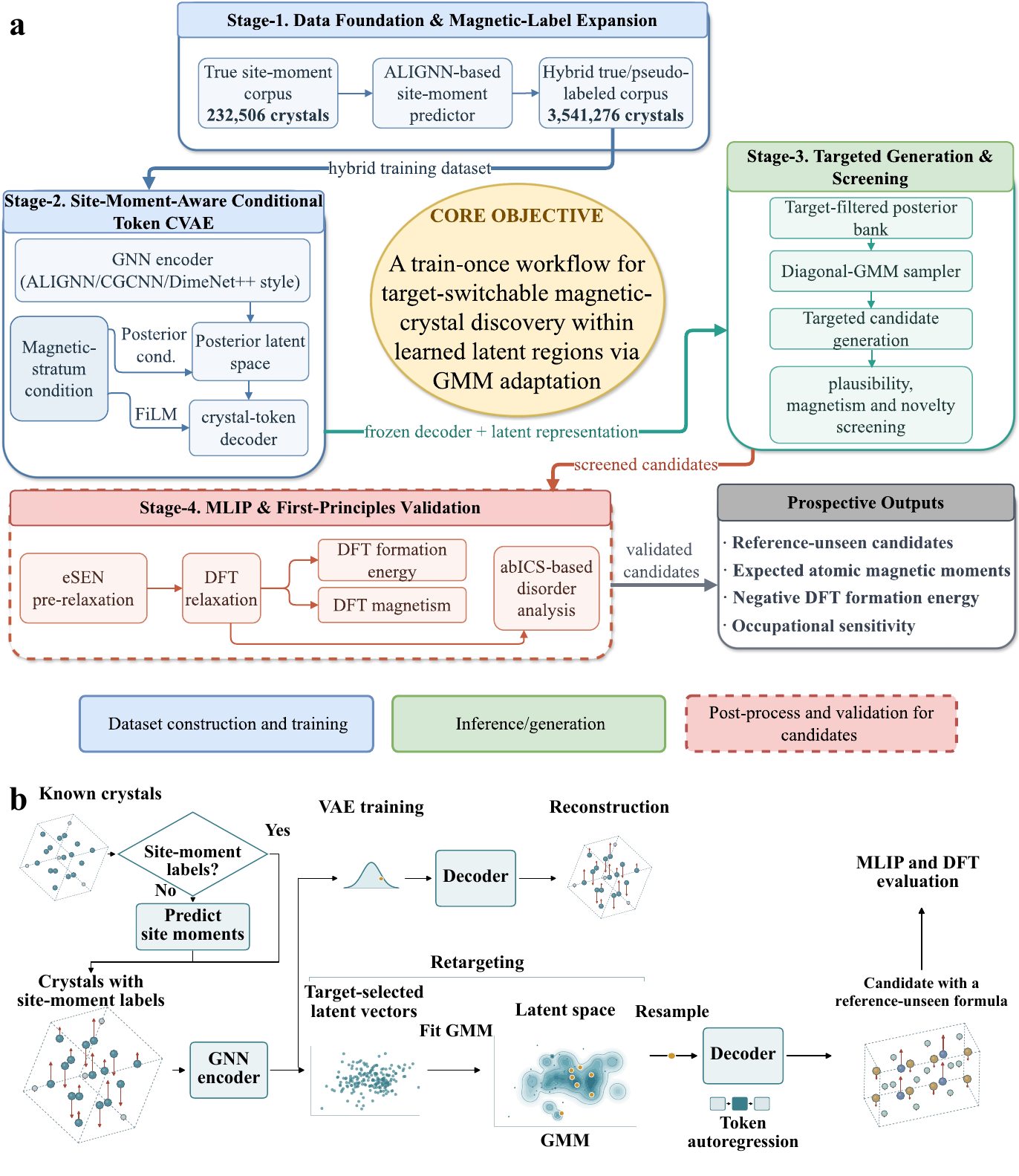}
\caption{GMMag workflow and core concept. a, Detailed workflow. Crystals with DFT-derived site magnetic moments train an ALIGNN predictor, which supplies magnetic labels only where these are absent. The combined dataset trains the generator. With generator weights fixed, selected reference structures supply latent vectors for replaceable GMM samplers. Candidates undergo geometry and predicted-magnetism screening, pre-relaxation with eSEN (an MLIP), and DFT evaluation; selected Nd--Fe--Co systems also undergo atomic site-occupancy sampling. ``Unseen'' denotes a reduced chemical formula absent from the fixed reference set. b, Core concept. The CVAE design includes posterior sampling for reconstruction, whereas production training predominantly used posterior means (Methods); the encoder means of selected reference structures supply a GMM fitted in 128 dimensions to redirect generation. The shared decoder predicts crystal attributes sequentially. The density shows the GMM's two-dimensional marginal. Crystal depictions illustrate workflow stages rather than verified reconstructions or discoveries; arrow lengths are schematic.}
\label{fig:workflow}
\end{figure}

GMMag links local magnetic-moment prediction, generator training, target-specific sampling and physical evaluation (Fig.~\ref{fig:workflow}a). An atomistic line graph neural network (ALIGNN) predictor is trained and evaluated on fixed splits of 232,506 structures with signed collinear magnetic moments at atomic sites, then supplies predicted labels for structures lacking them. DFT-derived labels take precedence; label source (DFT or prediction) determines sampling probabilities and magnetic-loss weights. The combined generator dataset contains 3,541,276 structures from the Materials Project (MP) \cite{jain2013materialsproject}, Novel Materials Discovery (NOMAD) \cite{draxl2018nomad}, Open Quantum Materials Database (OQMD) \cite{kirklin2015oqmd} and in-house calculations, including 174 in-house Nd-containing structures that enlarge the reference set available for Nd-targeted generation (Methods).

\begin{figure}[H]
\centering
\includegraphics[width=\textwidth,pagebox=cropbox]{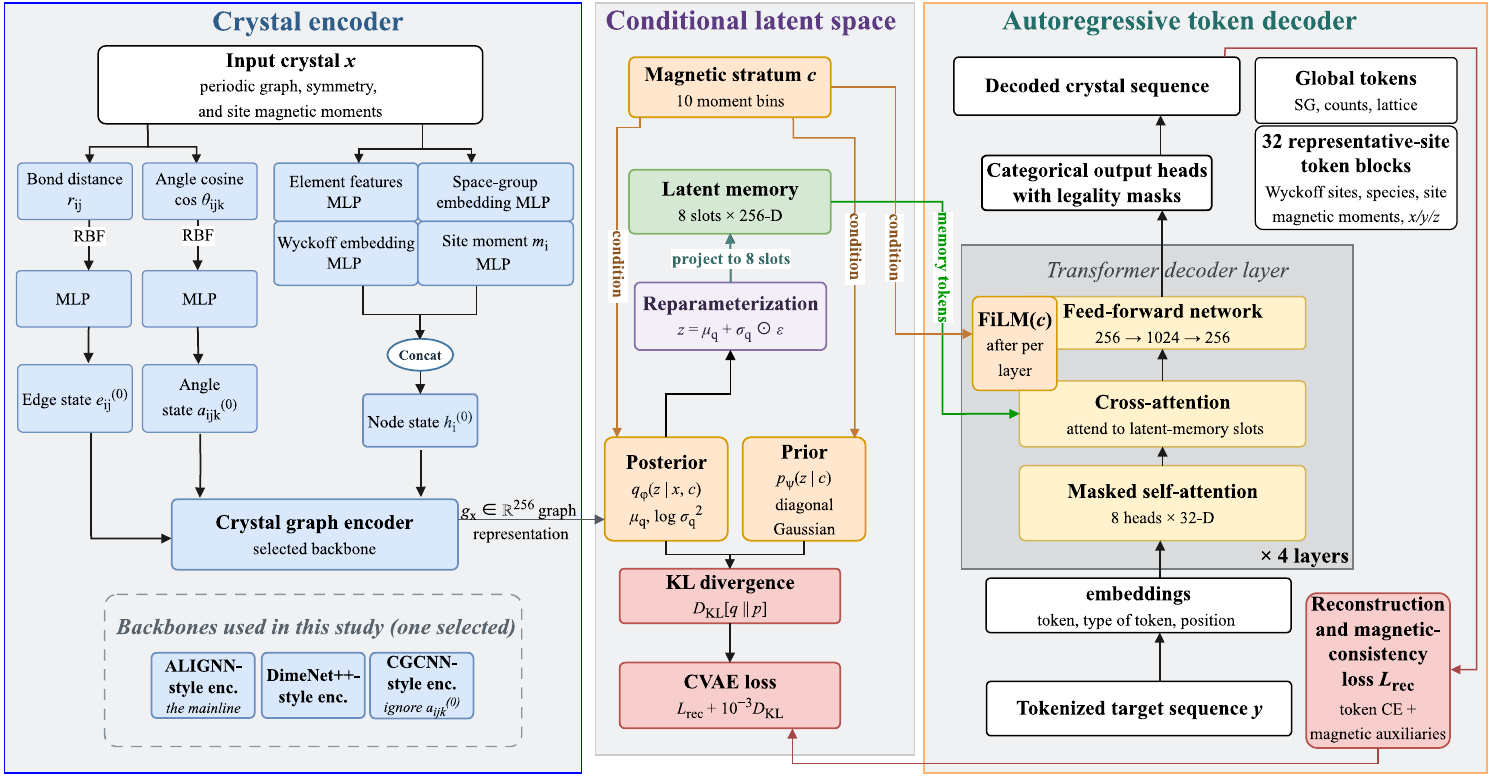}
\caption{Architecture of the CVAE generator with site-resolved magnetic moments used in GMMag. Independent feature branches feed the graph encoder; the magnetic condition enters the encoder's latent distribution (posterior), the learned conditional prior and the decoder. The decoder sequentially generates crystal-level descriptors and Wyckoff-site groups with signed mean magnetic moments. At generation time, its weights remain fixed and it takes GMM-sampled latent vectors and the target magnetic-condition bin. ALIGNN is the main encoder; the crystal graph convolutional neural network (CGCNN) and DimeNet++ \cite{gasteiger2020dimenetpp} are alternative encoders with the same latent and decoder interfaces; Supplementary Fig.~\suppref{SI-figS:encoder-sweep} compares their Nd-targeted generation. MLP, FiLM and GELU denote multilayer perceptron, feature-wise linear modulation \cite{perez2018film} and Gaussian error linear unit, respectively. Methods define the token encoding and complete training objective.}
\label{fig:model_architecture}
\end{figure}

The conditional variational autoencoder (CVAE) \cite{kingma2014vae,sohn2015cvae} uses separate encoder branches for elemental, space-group, Wyckoff-site-group and signed magnetic-moment features (Figs.~\ref{fig:workflow}b and~\ref{fig:model_architecture}). In this collinear representation, opposite signs denote oppositely oriented local magnetic moments along a common spin axis. An autoregressive decoder predicts discrete crystal descriptors (tokens) sequentially from preceding descriptors, the latent vector and a magnetic condition. Each Wyckoff-site group represents crystallographically equivalent sites by one signed mean magnetic moment. Reversing all moments together fixes a common sign convention while preserving relative orientations; the magnitude of their net sum per atom defines a ten-bin global magnetic condition. GMMag retains both moment magnitudes and signs. Its reconstruction objective includes additional losses for magnetic-moment signs, signed values and structure-level magnetic summaries (Eq.~\eqref{eq:cvae-loss}). A single moment per Wyckoff-site group cannot represent magnetic symmetry breaking within that group.

\FloatBarrier

In the workflow shown in Fig.~\ref{fig:workflow}a, the Stage-1 ALIGNN predictor used to supply missing site-resolved magnetic-moment labels achieved a local magnetic-moment mean absolute error (MAE) of $0.05~\mu_B$ on held-out labelled structures. The production checkpoint of the Stage-2 CVAE generator (architecture shown in Fig.~\ref{fig:model_architecture}) was evaluated separately by teacher-forced reconstruction of 4,096 test structures, yielding 95.8\% structural validity, 98.3\% composition matching and an MAE of $0.01~\mu_B$ for the reconstructed mean atomic magnetic moments within Wyckoff-site groups (Supplementary Table~\suppref{SI-tabS:stage1}).

The transformer decoder combines preceding token information through causal self-attention. Cross-attention lets its token representations read memory vectors projected from the CVAE latent vector $z$ \cite{vaswani2017attention}. Feature-wise linear modulation (FiLM) uses the global magnetic-condition bin to scale and shift each feature of the decoder hidden representations after every layer \cite{perez2018film}. These representations encode the evolving token sequence; FiLM does not directly modify $z$. Supplementary Note~\suppref{SI-sec:supp-decoder-operations} gives the details for both conditioning paths.

For the three evaluated checkpoints, ALIGNN and CGCNN produce similar numbers of distinct qualifying reference-unseen formulas, both exceeding DimeNet++, whereas DimeNet++ gives more distinct unseen formulas before screening (Supplementary Fig.~\suppref{SI-figS:encoder-sweep}). Section~\ref{sec:aligned-controls} tests decoder design and magnetic information during generation from the model's own preceding outputs.

\begin{figure}[!t]
\centering
\includegraphics[width=0.96\textwidth]{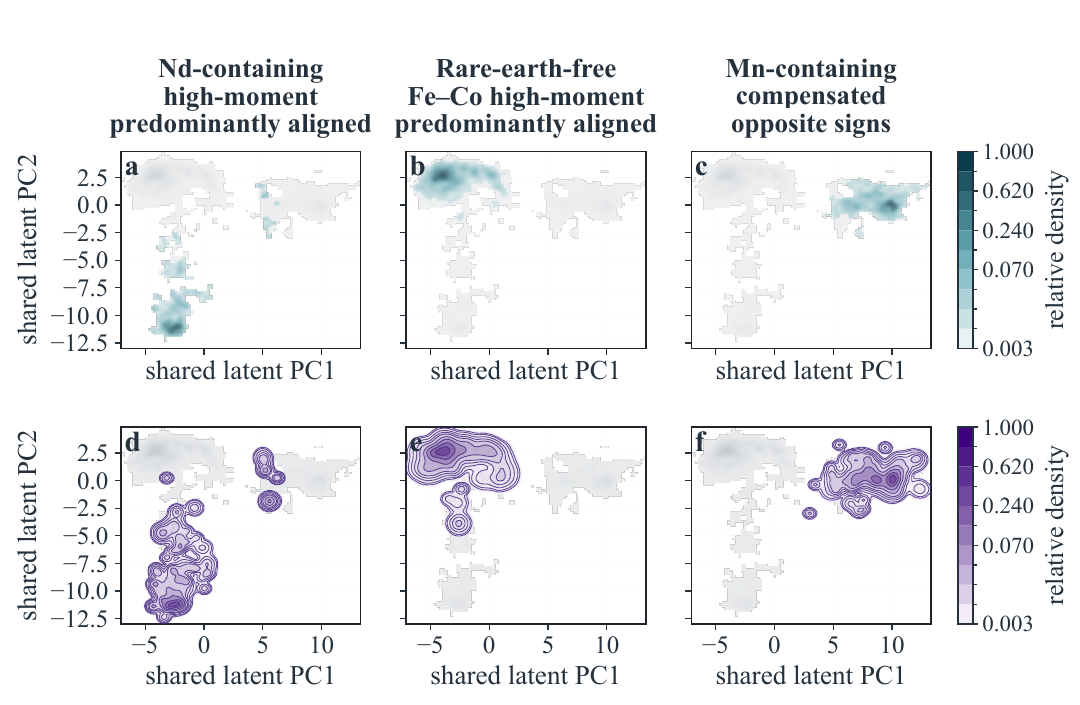}
\caption{Two-dimensional views of known-structure latent banks and their fitted distributions. a--c, Posterior means of known training structures selected for the Nd-containing high-net-moment, rare-earth-free Fe--Co high-net-moment and Mn-containing compensated-moment targets. d--f, Samples from diagonal GMMs fitted in the standardized 128-dimensional space ($K=64$, $\alpha=1.0$). All panels share a two-dimensional principal-component projection and axis limits; grey shading shows the pooled bank. Smoothed densities are normalized within each panel to compare spatial distributions, not absolute density magnitudes.}
\label{fig:samplers}
\end{figure}

\subsection{Adapting one trained generator to different targets}\label{sec:retargeting-results}

\begin{samepage}
We tested three generation tasks: Nd-containing and rare-earth-free Fe--Co-containing crystals with high net magnetic moments, and Mn-containing crystals with large, mutually compensating local magnetic moments. For $N$ atoms with signed local magnetic moments $m_i$, define the net-to-absolute moment ratio
\begin{equation}
r_{\rm align}=\begin{cases}m_{\rm net}/m_{\rm abs},&m_{\rm abs}>0,\\0,&m_{\rm abs}=0,\end{cases}\label{eq:magnetic-alignment}
\end{equation}
where $m_{\rm net}=|\sum_i m_i|/N$ is the magnitude of the net site-moment sum per atom and $m_{\rm abs}=\sum_i|m_i|/N$ is the mean local-moment magnitude. For the retargeting experiments in this subsection, we required $m_{\rm net}\geq1.8~\mu_B$/atom and $r_{\rm align}\geq0.9$ for the high-moment targets. For the compensated target in these experiments, we required $m_{\rm abs}\geq1.5~\mu_B$/atom, $m_{\rm net}\leq0.1~\mu_B$/atom, and local magnetic moments of both signs beyond $\pm0.1~\mu_B$.
\end{samepage}

For each task, we select known structures from the training partition of the combined dataset. The trained encoder maps each structure to a 128-dimensional distribution; its mean provides the structure's latent vector. These posterior-mean vectors form the latent bank, a reference set for subsequent exploration. Each latent coordinate is standardized before fitting a GMM with diagonal component covariances. With fitted weights $\widehat\pi_k$, means $\widehat\mu_k$, standard deviations $\widehat\sigma_k$, and bank mean and scale vectors $\boldsymbol\mu_{\rm bank}$ and $\boldsymbol s_{\rm std}$, sampling uses
\begin{align}
\Pr(k=j)&=\widehat\pi_j,\label{eq:gmm-component}\\*
u&=\widehat\mu_k+\alpha\widehat\sigma_k\odot\epsilon,\label{eq:gmm-standardized-sampling}\\*
z&=\boldsymbol\mu_{\rm bank}+\boldsymbol s_{\rm std}\odot u,\label{eq:gmm-sampling}
\end{align}
where $\Pr(k=j)$ denotes the probability of selecting mixture component $j$ ($j=1,\ldots,K$), $K$ is the number of mixture components, $\epsilon\sim\mathcal N(0,I)$ is a standard normal noise vector, and $\odot$ denotes element-wise multiplication. The scale $\alpha$ controls component spread. Reference structures are selected using stored magnetic labels; generated crystals are evaluated by the fixed ALIGNN predictor. A generated candidate qualifies if it passes decoding and geometry checks and meets the chemical and magnetic criteria specified for its task in that experiment; the screening pass rate is the fraction of all attempts meeting these criteria, before novelty filtering. Separately, reference-unseen means that its reduced formula is absent from the fixed formula reference compiled from all generator data splits, not just the latent bank. The count $D$ includes each distinct qualifying reference-unseen formula once per run.

The selected banks occupy different latent regions (Fig.~\ref{fig:samplers}). Both joint samplers use banks selected by chemistry and magnetism. Compared with Joint GMM, the single joint Gaussian yields more distinct qualifying reference-unseen formulas for Fe--Co but lower screening pass rates for all three targets, consistent with a unimodal fit smoothing over heterogeneous target regions. Joint GMM represents multiple local regions, balancing target qualification with formula exploration without maximizing either metric for every target.

\begin{table}[tbp]
\centering
\caption{Targeted generation with one fixed GMMag model. Screening pass rate is the percentage of generation attempts passing decoding, geometry, target chemistry and magnetic criteria, before formula-novelty filtering; $D$ counts distinct qualifying reference-unseen reduced formulas once per run. Qualifying means passing those task-specific screening criteria; reference-unseen means absent from the fixed 804,461-formula reference set. Values are means $\pm$ sample standard deviations over five seeds of 2,048 attempts, including invalid outputs. All runs use the dimensionless decoder temperature $\tau=0.8$ and top-$k=16$; bank Gaussians use spread scale $\alpha=1.0$. GMMs have $K=64$ and the joint Gaussian has $K=1$. Chemistry GMM and Magnetic GMM select reference structures by composition or magnetism alone; Joint Gaussian and Joint GMM use both criteria. Complete screening rates accompany the Source Data; additional sampling controls are described in Supplementary Note~\suppref{SI-sec:supp-retargeting}.}
\label{tab:retargeting}
\small
\setlength{\tabcolsep}{2.5pt}
\begin{tabular*}{\textwidth}{@{}l@{\extracolsep{\fill}}rrrrrr@{}}
\toprule
& \multicolumn{2}{c}{Nd target} & \multicolumn{2}{c}{Fe--Co target} & \multicolumn{2}{c}{Mn target}\\
\cmidrule(lr){2-3}\cmidrule(lr){4-5}\cmidrule(l){6-7}
Sampler & Pass (\%) & $D$ & Pass (\%) & $D$ & Pass (\%) & $D$\\
\midrule
Standard normal & \mbox{0.00 $\pm$ 0.00} & \mbox{0.0 $\pm$ 0.0} & \mbox{0.02 $\pm$ 0.03} & \mbox{0.0 $\pm$ 0.0} & \mbox{0.00 $\pm$ 0.00} & \mbox{0.0 $\pm$ 0.0} \\
Conditional prior & \mbox{0.01 $\pm$ 0.02} & \mbox{0.0 $\pm$ 0.0} & \mbox{0.04 $\pm$ 0.02} & \mbox{0.2 $\pm$ 0.4} & \mbox{0.00 $\pm$ 0.00} & \mbox{0.0 $\pm$ 0.0} \\
Chemistry GMM & \mbox{0.03 $\pm$ 0.03} & \mbox{0.0 $\pm$ 0.0} & \mbox{7.02 $\pm$ 0.58} & \mbox{3.6 $\pm$ 2.3} & \mbox{0.00 $\pm$ 0.00} & \mbox{0.0 $\pm$ 0.0} \\
Magnetic GMM & \mbox{0.11 $\pm$ 0.09} & \mbox{0.0 $\pm$ 0.0} & \mbox{0.89 $\pm$ 0.21} & \mbox{1.0 $\pm$ 0.7} & \mbox{2.89 $\pm$ 0.59} & \mbox{4.6 $\pm$ 1.5} \\
Joint Gaussian & \mbox{1.12 $\pm$ 0.17} & \mbox{11.6 $\pm$ 3.2} & \mbox{8.42 $\pm$ 0.57} & \mbox{12.2 $\pm$ 3.4} & \mbox{0.10 $\pm$ 0.06} & \mbox{0.4 $\pm$ 0.5} \\
Joint GMM & \mbox{15.48 $\pm$ 0.64} & \mbox{15.2 $\pm$ 1.3} & \mbox{18.72 $\pm$ 0.51} & \mbox{7.8 $\pm$ 2.3} & \mbox{3.63 $\pm$ 0.29} & \mbox{4.4 $\pm$ 1.7} \\
\bottomrule
\relax
\end{tabular*}
\end{table}

Changing the fitted sampling distribution therefore increases the frequency of desired outputs for targets represented in the reference structures. However, screening success alone does not determine how many distinct qualifying reference-unseen formulas a fixed number of generation attempts will yield.

\FloatBarrier
\subsection{From qualifying reference-unseen draws to distinct formulas}\label{sec:yield-results}

This section uses large-scale Nd-targeted generation to model how distinct qualifying reference-unseen formulas accumulate as the number of generation attempts increases.

To promote diversity, we expand the latent bank from 144 to 394 known structures by selecting the two highest magnetic-condition bins (lower boundary approximately $1.2~\mu_B$/atom) and omitting the net-to-absolute moment-ratio requirement (Methods). Generated candidates retain the $1.8~\mu_B$/atom ALIGNN-predicted threshold without a moment-ratio requirement. One draw denotes one generation attempt, including unsuccessful decoding. For draw $j$, let $H_j=1$ if it passes decoding, geometry, Nd chemistry and magnetic screening, and zero otherwise. Let $U_j=1$ if its reduced formula $F_j$ is absent from the fixed reference set $\mathcal R$, and zero otherwise; undecodable attempts have $H_j=U_j=0$. The joint indicator $I_j=H_jU_j$ selects qualifying reference-unseen draws. Their repeat-inclusive and distinct-formula counts are
\begin{align}
A_n&=\sum_{j=1}^{n}I_j=\sum_{j=1}^{n}H_jU_j,\label{eq:yield-counts}\\*
D_n&=\left|\{F_j:I_j=1,\ 1\leq j\leq n\}\right|.\label{eq:distinct-count}
\end{align}
Thus $A_n$ counts qualifying reference-unseen draws, including repeats, whereas $D_n$ counts their distinct reduced formulas.

For a single generation attempt, $Q=\Pr(I_j=1)=\Pr(H_j=1,U_j=1)$ is the probability of obtaining a qualifying reference-unseen candidate, before deduplication. It differs from the screening-pass probability $\Pr(H_j=1)$. Assuming that $Q$ remains approximately constant throughout a run, $\mathbb E[A_n]=nQ$. For independent qualifying reference-unseen draws with fixed conditional formula probabilities $p_f=\Pr(F_j=f\mid I_j=1)$, the exact expected distinct count is $\mathbb E[D\mid A=a]=\sum_f[1-(1-p_f)^a]$. This is a statistical occupancy model: each formula is counted once when first encountered. Repeated draws of common formulas reduce the number of distinct formulas per qualifying reference-unseen draw. Short generation runs cannot reliably estimate all $p_f$, motivating a low-parameter approximation to the accumulation curve.

For fitted GMM parameters $\widehat\Theta_K=\{\widehat\pi_k,\widehat\mu_k,\widehat\Sigma_k\}_{k=1}^{K}$ in $d=128$ standardized latent dimensions, define
\begin{equation}
\mathcal C(\widehat\Theta_K)=\frac{1}{d}\sum_{k=1}^{K}\widehat\pi_k
\left[\ln\widehat\pi_k-\frac{1}{2}\ln|\widehat\Sigma_k|\right].
\label{eq:gmm-descriptor}
\end{equation}
This descriptor summarizes component weights and covariance volumes before scaling component spread. Its three calibration values correspond one-to-one to $K$, so their effects cannot be separated. Writing $C=\mathcal C(\widehat\Theta_K)$, and using the sampler identity $\rho_{\rm rms}=\alpha$ derived in Methods, the response for this joint probability is
\begin{equation}
\operatorname{logit}\widehat Q=b_0+b_\alpha\alpha+b_\tau\tau+b_CC+b_{\alpha C}\alpha C.
\label{eq:qualifying-response}
\end{equation}
The logit link enforces probability bounds, while $\alpha C$ allows the association between latent sampling spread and the joint log-odds to vary among fitted mixtures.

We approximate distinct-formula accumulation with a smooth crossover from nearly linear accumulation towards a finite-window exponent $0<\beta_{\rm late}<1$:
\begin{equation}
\mathcal D(a;\beta_{\rm late},a_0,\gamma)=
a\left[\frac{1+(a/a_0)^\gamma}{1+(1/a_0)^\gamma}\right]^{-(1-\beta_{\rm late})/\gamma},
\qquad \mathcal D(0)=0.
\label{eq:occupancy-law}
\end{equation}
Normalization gives $\mathcal D(1)=1$. The local logarithmic slope,
\begin{equation}
\frac{\partial\ln\mathcal D}{\partial\ln a}
=1-(1-\beta_{\rm late})\frac{(a/a_0)^\gamma}{1+(a/a_0)^\gamma},
\label{eq:occupancy-slope}
\end{equation}
decreases from near one towards $\beta_{\rm late}$ and equals $(1+\beta_{\rm late})/2$ at $a=a_0$. Thus $a_0$ measures the crossover in qualifying reference-unseen draws, including repeats, and $\gamma>0$ controls its sharpness. This exponent describes the fitted range of $A_n$; a finite formula set would ultimately saturate.

The accumulation parameters use the responses
\begin{align}
\operatorname{logit}\widehat\beta_{\rm late}&=c_0+c_\tau\tau,\label{eq:late-exponent-response}\\*
\ln\widehat a_0&=d_0+d_\alpha\alpha+d_\tau\tau+d_CC+d_{\alpha C}\alpha C,\label{eq:crossover-response}\\*
\ln\widehat\gamma&=g_0.\label{eq:sharpness-response}
\end{align}
The links enforce $0<\widehat\beta_{\rm late}<1$ and positive $\widehat a_0,\widehat\gamma$. We retain a decoder-temperature-only response for $\beta_{\rm late}$ as a parsimonious choice supported by the comparison of candidate response models on the calibration grid (Supplementary Table~\suppref{SI-tabS:occupancy-selection}). The crossover response allows both latent spread and decoder temperature to contribute, with effects of latent spread that depend on the fitted mixture.

Under an independent Bernoulli working model, marginalizing over $A_n\sim\operatorname{Binomial}(n,\widehat Q)$ gives the prediction for $n$ generation attempts
\begin{equation}
\widehat D_n=\sum_{a=0}^{n}\mathcal D(a;\widehat\beta_{\rm late},\widehat a_0,\widehat\gamma)
\binom{n}{a}\widehat Q^a(1-\widehat Q)^{n-a}.
\label{eq:composed-yield}
\end{equation}
Calibration used 45 sampling-parameter settings with 50,000 draws each. We fit $Q$ from the repeat-inclusive counts $A_n$ at each setting, then fit the accumulation-response coefficients jointly using observed $(A_n,D_n)$ curves. The two fitted responses are combined through Eq.~\eqref{eq:composed-yield}; fitting objectives and optimization details are given in Methods. Table~\ref{tab:scaling-coefficients} reports the coefficients refitted on all 45 settings after validation.

\begin{table}[!htbp]
\caption{Coefficients for predicting distinct-formula yield at a fixed generation budget. The response for the joint probability $Q$ follows Eq.~\eqref{eq:qualifying-response}; the remaining rows parameterize the accumulation responses above. Here $C=\mathcal C(\widehat\Theta_K)$ and $\rho_{\rm rms}=\alpha$. Coefficients use the original covariate scales; dashes denote omitted terms.}
\label{tab:scaling-coefficients}
\centering
\small
\setlength{\tabcolsep}{3pt}
\begin{tabular*}{\textwidth}{@{}l@{\extracolsep{\fill}}rrrrr@{}}
\toprule
Response & Intercept & $\alpha$ & $\tau$ & $C$ & $\alpha C$\\
\midrule
$\operatorname{logit}\widehat Q$ & $b_0=-2.597$ & $b_\alpha=-1.733$ & $b_\tau=-0.664$ & $b_C=-0.455$ & $b_{\alpha C}=0.738$ \\
$\operatorname{logit}\widehat\beta_{\rm late}$ & $c_0=-0.778$ & -- & $c_\tau=1.183$ & -- & -- \\
$\ln\widehat a_0$ & $d_0=0.807$ & $d_\alpha=2.329$ & $d_\tau=0.022$ & $d_C=-0.027$ & $d_{\alpha C}=-0.781$ \\
$\ln\widehat\gamma$ & $g_0=0.397$ & -- & -- & -- & -- \\
\bottomrule
\relax
\end{tabular*}
\end{table}

\begin{figure}[tbp]
\centering
\includegraphics[width=\textwidth]{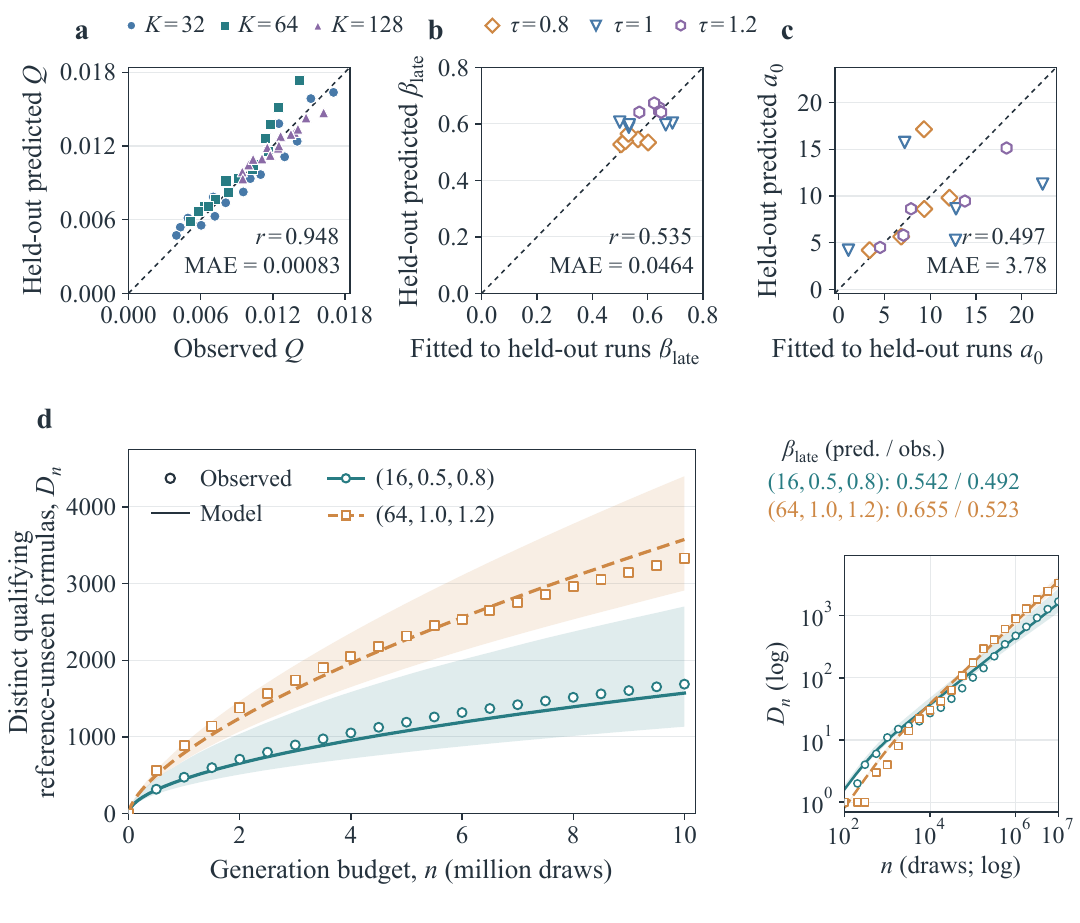}
\caption{Predicting distinct-formula yield at a fixed generation budget. a, Observed and cross-validated probability $Q$ of a qualifying reference-unseen draw for the 45 sampling-parameter settings. b,c, Held-out predictions of $\beta_{\rm late}$ and $a_0$ against estimates pooled over the three $K$ values for each held-out $(\alpha,\tau)$ combination. Shapes denote decoder temperature; dashed diagonals indicate equality. d, Observed and predicted $D_n$ for two independent ten-million-draw runs, labelled by $(K,\alpha,\tau)$; open symbols are observations and shading denotes sensitivity ranges obtained by resampling parameter combinations, not confidence intervals. The inset shows logarithmic axes.}
\label{fig:scaling-model}
\end{figure}

Five-fold cross-validation holding out complete sampling-parameter settings captured the principal variation in $Q$. Prediction of the accumulation parameters was less accurate when all three $K$ values for each held-out $(\alpha,\tau)$ combination were withheld together (Fig.~\ref{fig:scaling-model}a--c). These tests assess interpolation within the recorded parameter grid.

The coefficients describe how sampling affects $Q$ and formula repetition. Across the three calibrated mixtures, $\partial_\alpha\operatorname{logit}\widehat Q=b_\alpha+b_{\alpha C}C<0$, whereas $\partial_\alpha\ln\widehat a_0=d_\alpha+d_{\alpha C}C>0$. Narrower latent sampling therefore increases the fitted joint probability $Q$ while bringing the repetition-limited crossover forward. Higher decoder temperature lowers $\widehat Q$ ($b_\tau<0$) but raises the late-window exponent ($c_\tau>0$), from 0.542 at $\tau=0.8$ to 0.655 at $\tau=1.2$. These competing associations explain why $Q$ alone does not determine accumulated discovery; their net effect depends on the requested budget through Eq.~\eqref{eq:composed-yield}.

The long runs test the fitted response beyond its calibration budget (Fig.~\ref{fig:scaling-model}d). The endpoint at $(K,\alpha,\tau)=(64,1.0,1.2)$, within the calibrated parameter range, agrees with the fitted model prediction. The result at $(16,0.5,0.8)$, outside the calibrated component counts, reflects compensation between errors in $Q$ and formula accumulation. Neither run was used to fit the coefficients. This relation supports planning for the calibrated workflow; changing the generator, reference structures or screening rules requires recalibration.
\FloatBarrier

\subsection{Effects of model architecture and magnetic information on generation}\label{sec:aligned-controls}

The distinction between screening success and distinct-formula accumulation also informs model design. We therefore use a series of ablation studies to examine how decoder architecture, conditioning paths, and the representation and supervision of site-resolved magnetic moments affect model performance. We revisit the Nd, Fe--Co and Mn chemical targets of Section~\ref{sec:retargeting-results}, retaining their net-moment thresholds but omitting the additional alignment and local-moment constraints (Table~\ref{tab:aligned-controls}). The baseline uses the mainline configuration, retrained separately from the production checkpoint. Repeated training and generation runs assess run-to-run variability.

\begin{figure}[!htbp]
\centering
\includegraphics[width=\textwidth]{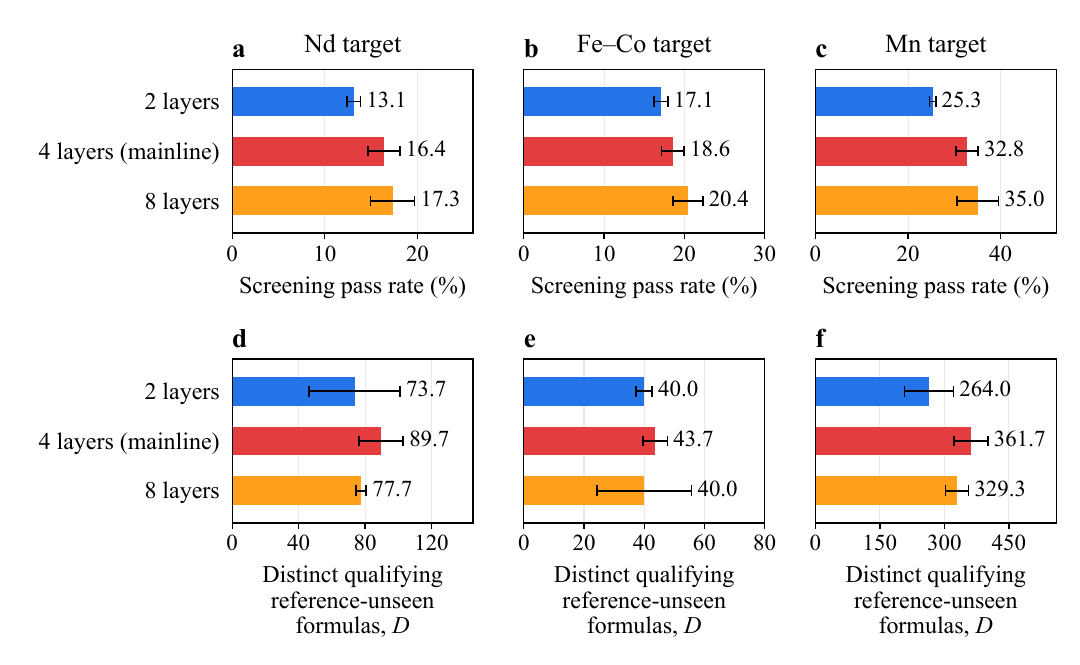}
\caption{Depth of the transformer decoder and targeted generation. Columns correspond to the Nd, Fe--Co and Mn targets. a--c, Screening pass rate. d--f, Number $D$ of distinct qualifying reference-unseen formulas. Bars show means and error bars show sample standard deviations across three independently trained models; numbers give means. Blue, red and orange denote two, four and eight layers. The four-layer mainline architecture is evaluated using Signed-CE\&Aux controls trained separately from the production checkpoint. Each model uses one shared generation seed and 16,384 attempts per target, including invalid outputs. Screening uses the criteria in Table~\ref{tab:aligned-controls}, before formula-novelty filtering against the fixed 804,461-formula reference set. Supplementary Note~\suppref{SI-sec:supp-decoder-capacity} gives parameter counts, reconstruction metrics and the matched-update protocol.}
\label{fig:decoder-depth}
\end{figure}

We first examine decoder capacity. Increasing the transformer decoder from two to four to eight layers reduces held-out token cross-entropy from 0.679 to 0.637 to 0.605 and lowers errors in lattice length $a$ and Wyckoff parameters under teacher forcing (Supplementary Table~\suppref{SI-tabS:decoder-capacity}). Mean screening pass rates also increase with depth, but distinct qualifying reference-unseen formula counts do not (Fig.~\ref{fig:decoder-depth}). At 16,384 attempts per trained model and target, four layers give the highest mean count for all three targets among the tested depths, supporting the capacity chosen for GMMag. Screening-passing eight-layer outputs contain a larger fraction of reference-known formulas than four-layer outputs (Supplementary Note~\suppref{SI-sec:supp-decoder-capacity}).

\begin{figure}[!htbp]
\centering
\includegraphics[width=\textwidth]{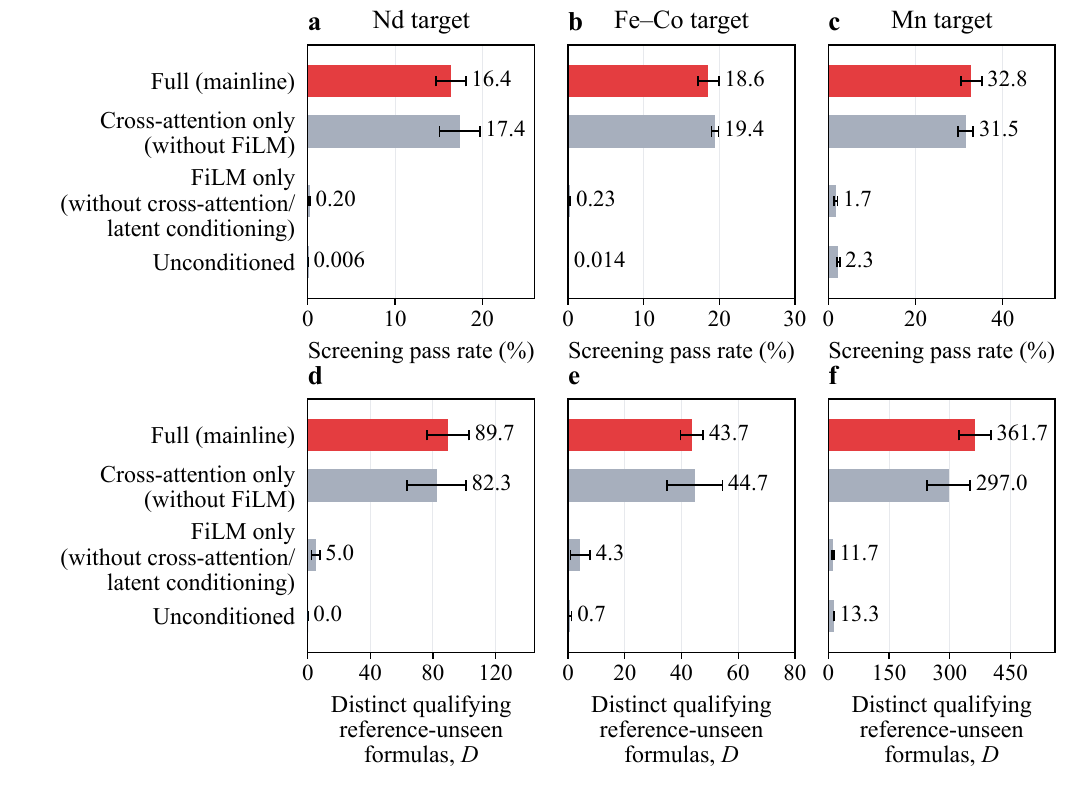}
\caption{Decoder conditioning paths and targeted generation. Columns correspond to the Nd, Fe--Co and Mn targets. a--c, Screening pass rate. d--f, Number $D$ of distinct qualifying reference-unseen formulas. All models retain four decoder layers and causal self-attention; ``only'' refers to conditioning paths. Full (mainline) retains both paths and reuses the four-layer baseline in Fig.~\ref{fig:decoder-depth}. Cross-attention only removes FiLM; FiLM only removes the encoder--latent path and KL regularization; Unconditioned removes both paths. Bars and numbers show means, and error bars show sample standard deviations across three independently trained models, each evaluated with 16,384 attempts per target. The screening, reference-novelty and generation protocols match Fig.~\ref{fig:decoder-depth}. Supplementary Note~\suppref{SI-sec:supp-decoder-paths} details the conditioning controls.}
\label{fig:decoder-conditioning}
\end{figure}

For the four-layer baseline, teacher-forced reconstruction of 17,737 test structures gave $(97.62\pm0.12)$\% composition matching, a lattice-length $a$ MAE of $0.143\pm0.007$~\AA{} and a periodic fractional-coordinate MAE of $0.0853\pm0.0023$ over active Wyckoff parameters (mean $\pm$ sample standard deviation across three training seeds; Supplementary Table~\suppref{SI-tabS:decoder-capacity}).

We next compare the two conditioning paths introduced in Section~\ref{sec:site-moment-results}: cross-attention reads latent memory, whereas FiLM modulates decoder hidden representations using the magnetic-condition bin. Removing FiLM retains most of the distinct-formula yield, with target-dependent changes (Fig.~\ref{fig:decoder-conditioning}d--f). Removing the entire latent path, leaving only FiLM, sharply reduces targeted discovery; removing both paths gives an unconditioned autoregressive model with similarly low yields. The FiLM only control also removes the encoder, latent sampling and Kullback--Leibler (KL) regularization, so this comparison tests the complete latent-conditioned generation route rather than cross-attention alone. All four configurations retain signed magnetic-moment outputs and magnetic auxiliary losses (Supplementary Table~\suppref{SI-tabS:decoder-paths}).

\begin{table}[!htbp]
\centering
\caption{Generation under matched conditions and magnetic-moment sign agreement. a, $D_{100\mathrm{k}}$ counts distinct qualifying reference-unseen formulas per 100,000 attempts. Exact magnetic-moment sign agreement is evaluated for geometry- and chemistry-eligible structures with both positive and negative nonzero ALIGNN-predicted moments. All sites with nonzero predicted moments must match the generated signs after one whole-structure sign alignment chosen to minimize moment MAE. Sites span all constituent elements. Each eligible generated structure contributes equally within a run, and the three generation tasks receive equal weight. b, Screening pass rates precede formula-novelty filtering. ALIGNN predictions must satisfy $m_{\rm net}\geq1.8~\mu_B$/atom for Nd and Fe--Co, or $m_{\rm net}\leq0.1~\mu_B$/atom for Mn, including nonmagnetic structures. Invalid attempts remain in the denominators. Values are means $\pm$ sample standard deviations over three training seeds after averaging three generation seeds. A slash denotes an inapplicable local-moment output metric; bold identifies the highest mean without implying significance. Methods and Supplementary Note~\suppref{SI-sec:supp-aligned-controls} give the nonzero threshold and complete averaging procedure.}
\label{tab:aligned-controls}
\small
\setlength{\tabcolsep}{2.7pt}
\begin{tabular*}{\textwidth}{@{}l@{\extracolsep{\fill}}rrrr@{}}
\toprule
\multicolumn{5}{@{}l}{\textbf{a, Distinct formulas and magnetic-moment sign agreement}}\\
\midrule
Model & \shortstack{Nd target\\$D_{100\mathrm{k}}$} & \shortstack{Fe--Co target\\$D_{100\mathrm{k}}$} & \shortstack{Mn target\\$D_{100\mathrm{k}}$} & \shortstack{Exact magnetic-moment\\sign agreement (\%)}\\
\midrule
\begin{tabular}[t]{@{}l@{}}Signed-CE\&Aux\\(GMMag mainline)\end{tabular} & $\boldsymbol{273.8 \pm 71.0}$ & $165.2 \pm 9.2$ & $\boldsymbol{1964.4 \pm 195.5}$ & $\boldsymbol{37.27 \pm 0.74}$ \\
Signed-CE & $219.9 \pm 38.0$ & $164.9 \pm 8.8$ & $1883.0 \pm 225.9$ & $34.68 \pm 1.52$ \\
EncAbs-CE & $270.7 \pm 76.1$ & $152.1 \pm 26.4$ & $1766.9 \pm 143.1$ & $33.35 \pm 1.31$ \\
HistAbs-CE & $247.6 \pm 59.3$ & $161.1 \pm 32.4$ & $1737.8 \pm 75.6$ & $34.71 \pm 1.21$ \\
DualAbs-CE & $234.6 \pm 22.6$ & $139.8 \pm 2.8$ & $1567.8 \pm 215.3$ & $31.13 \pm 0.49$ \\
Global-GReg & $252.7 \pm 25.5$ & $\boldsymbol{165.3 \pm 21.1}$ & $1350.4 \pm 348.8$ & / \\
\bottomrule
\relax
\end{tabular*}

\medskip
\begin{tabular*}{\textwidth}{@{}l@{\extracolsep{\fill}}rrr@{}}
\toprule
\multicolumn{4}{@{}l}{\textbf{b, Screening pass rate per generation attempt (\%)}}\\
\midrule
Model & Nd target & Fe--Co target & Mn target\\
\midrule
Signed-CE\&Aux & $16.38 \pm 1.96$ & $18.53 \pm 1.87$ & $32.76 \pm 2.29$ \\
Signed-CE & $15.22 \pm 3.91$ & $18.62 \pm 0.61$ & $32.17 \pm 4.87$ \\
EncAbs-CE & $\boldsymbol{16.96 \pm 2.98}$ & $20.45 \pm 0.81$ & $\boldsymbol{33.63 \pm 5.81}$ \\
HistAbs-CE & $16.13 \pm 3.95$ & $18.93 \pm 3.41$ & $31.85 \pm 2.27$ \\
DualAbs-CE & $16.61 \pm 3.99$ & $21.44 \pm 2.57$ & $31.86 \pm 4.65$ \\
Global-GReg & $14.78 \pm 0.78$ & $\boldsymbol{22.29 \pm 1.27}$ & $29.92 \pm 3.24$ \\
\bottomrule
\relax
\end{tabular*}
\end{table}

Finally, we examine magnetic representation and supervision using six model configurations for ablation experiments, with three training seeds and three generation seeds per trained model and 100,000 attempts per task (Table~\ref{tab:aligned-controls}). Signed-CE\&Aux uses the complete GMMag objective in Eq.~\eqref{eq:cvae-loss}; Signed-CE removes its three additional magnetic losses. EncAbs-CE, HistAbs-CE and DualAbs-CE also replace signed magnetic-moment inputs with their absolute values in the encoder, preceding decoder inputs or both, while retaining signed output targets and magnetic-moment token cross-entropy (CE). Global-GReg instead uses the ten-bin global magnetic condition and predicts a continuous global moment, without local magnetic-moment inputs or outputs. Its architecture, objective and initialization differ, making it a comparison of complete systems. Each model separately encodes the same selected reference structures within each task.

GMMag (Signed-CE\&Aux) has the highest task-averaged exact magnetic-moment sign agreement and the highest distinct-formula yields for the Nd and Mn targets, although it does not maximize screening pass rate. Additional magnetic supervision improves sign agreement over Signed-CE by 2.6 percentage points, with gains in all three training seeds driven by Mn. The advantage narrows with equal formula weighting. With equal weight per generated occurrence, Signed-CE slightly trails HistAbs-CE in task-averaged sign agreement; the controls therefore establish no universal ordering (Supplementary Note~\suppref{SI-sec:supp-aligned-controls}). This metric measures agreement between generated magnetic-moment signs and ALIGNN predictions for the same structures; it does not measure magnetic-order accuracy against DFT.

For the Mn target, GMMag's larger reference-unseen fraction and lower formula repetition compensate for its lower screening pass rate relative to EncAbs-CE (Supplementary Fig.~\suppref{SI-figS:aligned-accumulation} and Table~\suppref{SI-tabS:aligned-quality}). Together, the architecture and magnetic-information comparisons show why model selection must consider distinct-formula discovery alongside reconstruction, screening success and magnetic-moment agreement.

\FloatBarrier
\relax

\subsection{Physical evaluation of generated candidates}\label{sec:prospective-validation}

For physical evaluation, a separate ten-million-draw Nd-targeted run of the same generator at $(K,\alpha,\tau)=(64,1.0,1.2)$ supplied candidates for pre-relaxation with the equivariant Smooth Energy Network (eSEN) \cite{fu_esen} and DFT. Of 3,320 distinct qualifying reference-unseen formulas, 42 candidates met the eSEN and surrogate selection criteria. The resulting set contains 38 candidates with accepted DFT results: 33 newly converged calculations and five previous calculations confirmed to represent equivalent structures. Among them, 34 have negative formation energy using matched elemental references, 26 exceed a corrected collinear magnetization estimate of $1.5$~T, and 22 meet both criteria. The screening steps and exclusions are retained in Supplementary Table~\suppref{SI-tabS:dft-funnel}.

\begin{figure}[htbp]
\centering
\includegraphics[width=\textwidth]{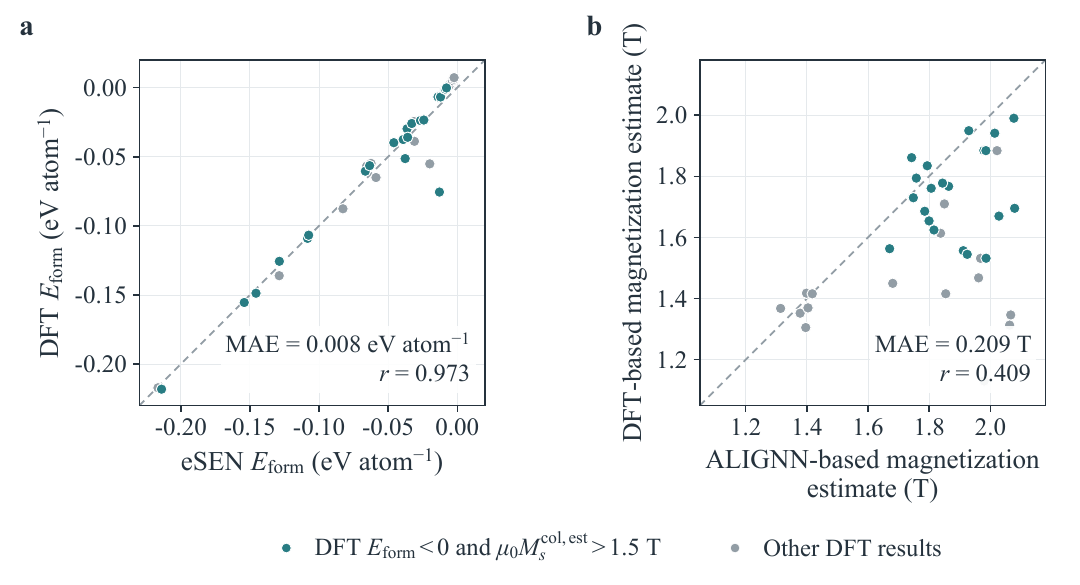}
\caption{Physical validation of generated Nd-containing candidates. a, eSEN versus DFT formation energies using matched elemental references for the 38 usable results from the illustrative Nd-targeted generation run. b, Screening estimates versus DFT-derived corrected collinear magnetization estimates for the same candidates. Dashed diagonals indicate equality; teal points meet both $E_{\rm form}^{\rm DFT}<0$ and $\mu_0M_s^{\rm col,est}>1.5$~T.}
\label{fig:prospective}
\end{figure}

The eSEN--DFT comparison supports energy-based screening within this selected set (Fig.~\ref{fig:prospective}a). Elemental reference energies use settings matched to the corresponding candidate calculations. The magnetic comparison tests screening predictions against DFT-derived collinear magnetization estimates with prescribed corrections for omitted rare-earth $4f$ moments (Fig.~\ref{fig:prospective}b). These estimates assume saturated sublattice alignment and do not establish a magnetic ground state.

Table~\ref{tab:representative-dft} highlights twelve reference-unseen candidates across eleven Nd-containing chemical systems, drawn from this run and other sampling-parameter settings. The complete 62-candidate collection, including positive formation energies, is retained in Supplementary Table~\suppref{SI-tabS:pooled-dft}. It includes candidates from the same run and therefore provides a broader collection, not an independent evaluation set.

\begin{table}[htbp]
\centering
\small
\caption{Representative DFT-evaluated generated candidates, ordered by decreasing corrected collinear magnetization estimate. These are selected entries from the full collection in Supplementary Table~\suppref{SI-tabS:pooled-dft}. Novelty is relative to the fixed generator formula reference. SG is the relaxed space-group number and $N$ the computational-cell atom count. Corrected collinear magnetization estimates and DFT formation energies refer to the relaxed candidates; energies above the convex hull are estimated separately with eSEN. Energies are rounded for display; selection uses unrounded values. Negative formation energy alone does not establish stability against competing phases.}
\label{tab:representative-dft}
\setlength{\tabcolsep}{2pt}
\begin{tabular}{@{}lrrrrr@{}}
\toprule
Formula & SG & $N$ & \shortstack{$\mu_0M_s^{\rm col,est}$\\(T)} & \shortstack{$E_{\rm form}^{\rm DFT}$\\(eV atom$^{-1}$)} & \shortstack{$E_{\rm hull}^{\rm eSEN}$\\(eV atom$^{-1}$)}\\
\midrule
Nd$_{2}$Fe$_{14}$Co$_{3}$ & 166 & 57 & 1.990 & -0.001 & 0.032 \\
NdFe$_{7}$Co$_{4}$ & 127 & 24 & 1.970 & -0.006 & 0.048 \\
NdSmFe$_{14}$Co$_{3}$ & 160 & 57 & 1.854 & -0.011 & 0.024 \\
NdYFe$_{14}$Co$_{3}$ & 160 & 57 & 1.838 & -0.024 & 0.023 \\
NdFe$_{8}$O & 1 & 20 & 1.814 & -0.269 & 1.025 \\
Nd(Fe$_{5}$P)$_{3}$ & 160 & 57 & 1.695 & -0.155 & 0.173 \\
Nd(Fe$_{5}$Si)$_{3}$ & 160 & 57 & 1.686 & -0.107 & 0.127 \\
NdSmVFe$_{16}$ & 160 & 57 & 1.654 & 0.000 & 0.025 \\
NdCr$_{2}$F & 8 & 8 & 1.625 & -1.026 & 2.018 \\
NdFe$_{11}$S$_{3}$ & 173 & 30 & 1.563 & -0.036 & 0.416 \\
NdZrFe$_{10}$ & 47 & 12 & 1.545 & -0.023 & 0.050 \\
NdFe$_{8}$N$_{3}$ & 1 & 24 & 1.532 & -0.051 & 0.251 \\
\multicolumn{6}{@{}l@{}}{\footnotesize An excerpt from the full candidate list.}\\
\bottomrule
\relax
\end{tabular}
\end{table}

\begin{figure}[htbp]
\centering
\includegraphics[width=0.98\textwidth]{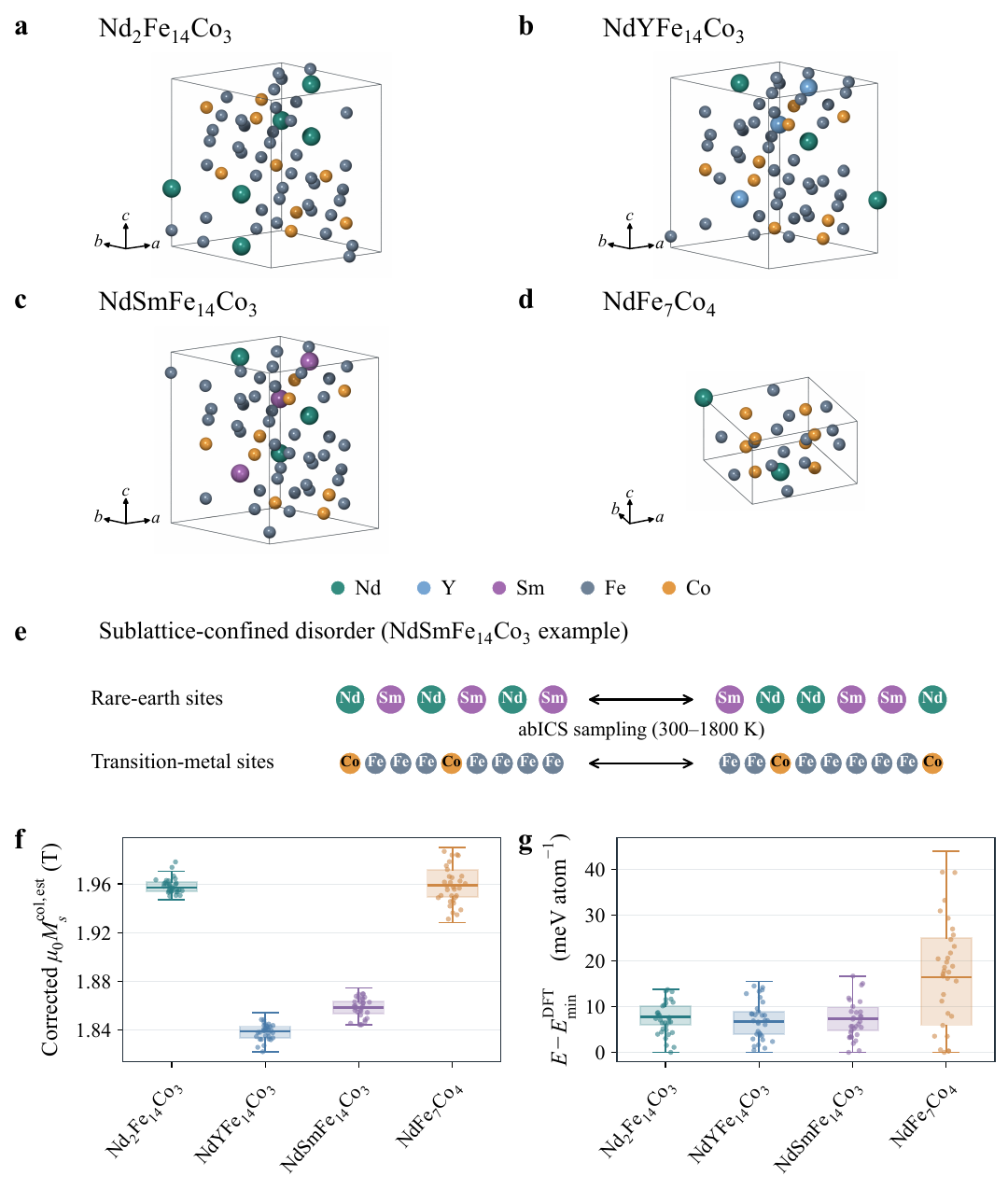}
\caption{Atomic site-occupancy disorder in four promising Nd--Fe--Co candidates. a--d, DFT-relaxed parent structures of four promising reference-unseen candidates. e, Schematic Nd/Sm and Fe/Co exchanges within their respective rare-earth and transition-metal sublattices at fixed composition; all six rare-earth sites and an illustrative subset of transition-metal sites are shown. f, DFT-derived collinear magnetization estimates including the prescribed open-core $4f$ correction. g, DFT energies relative to the lowest-energy DFT-evaluated configuration within each system. abICS used aenet-assisted replica-exchange sampling with ten temperature replicas spanning 300--1800~K at fixed composition and parent topology. Each box pools 110 DFT-evaluated configurations across the initial and active-learning rounds: centres are medians, boxes span the interquartile range and whiskers extend to observations within 1.5 times that range. Dots show a deterministic subsample of configurations.}
\label{fig:abics-disorder}
\end{figure}
\begin{samepage}
We used the ab initio Configuration Sampling Toolkit (abICS) \cite{kasamatsu2023abics} to examine atomic site-occupancy disorder in four promising Nd--Fe--Co candidates (Fig.~\ref{fig:abics-disorder}a--d). At fixed composition and parent topology, an on-lattice Atomic Energy Network (aenet) model \cite{artrith2016aenet} guided replica-exchange Monte Carlo over ten replicas spanning 300--1800~K, exploring Fe/Co and, where applicable, rare-earth exchanges within their respective sublattices (Fig.~\ref{fig:abics-disorder}e). Starting from 50 DFT-evaluated configurations, three active-learning rounds each added 20 DFT-evaluated configurations, giving 110 configurations per parent. The DFT-derived corrected collinear magnetization estimates remain within narrow, system-dependent ranges, whereas relative energies are more sensitive to occupation (Fig.~\ref{fig:abics-disorder}f,g). These pooled calculations probe sensitivity to atomic site occupations; they do not establish an equilibrium disorder distribution.
\par
\end{samepage}
\section{Discussion}

GMMag combines site-resolved magnetic information with replaceable sampling distributions. Modelling a fixed number of generation attempts shows how repetition limits the accumulation of different formulas, making screening pass rates and distinct-formula counts complementary measures of performance. The controlled experiments support a benefit of additional magnetic supervision for exact magnetic-moment sign agreement, with equal weight per eligible generated structure within each run, while formula-discovery gains depend on the generation target.

The decoder comparisons extend this distinction to model capacity: better teacher-forced prediction and higher screening pass rates need not yield more distinct qualifying reference-unseen formulas. The larger share of reference-known formulas among screening-passing eight-layer outputs motivates the hypothesis that increased capacity strengthens reliance on learned autoregressive regularities, potentially favouring familiar compositions over exploration through latent sampling. Related sequence VAEs show that autoregressive decoders can reduce their use of latent information \cite{bowman2016continuous}. Our results do not directly measure that dependence or establish memorization of individual training structures; the four-layer choice is an empirical balance within the tested settings.

The demonstrated target adaptation uses regions represented by known training structures. Formula novelty is evaluated against the fixed reference set, and magnetic-moment agreement against the fixed ALIGNN predictor. First-principles calculations add independent physical evidence for selected structures, but the single-initialization collinear protocol leaves competing magnetic orders and spin--orbit effects unresolved. Extending those calculations, testing additional targets represented in the training data and validating the yield response under new workflows would establish how broadly the demonstrated retargeting strategy transfers.
\FloatBarrier
\relax

\newpage
\section{Methods}\label{sec:methods}

\subsection{Data, site-resolved magnetic moments and sign conventions}\label{sec:data-gauge}

DFT-derived labels combine AFLOW and OQMD records with in-house Nd-containing calculations using the Vienna Ab initio Simulation Package (VASP) \cite{kresse1996vasp,kresse1996efficiency} and Perdew--Burke--Ernzerhof (PBE) functional \cite{perdew1996pbe}. The retained public records are spin-polarized and collinear. Each $m_i$ is a signed atom-projected spin moment along the common axis; because the sources lack a common projection partition, site-moment sums are learning and screening aggregates distinct from the complete VASP cell spin moment used in physical evaluation.

Geometry and site-moment arrays are checked for consistent counts and ordering; in-house pairs come from the same completed calculation. Non-finite arrays, inconsistent lengths and graph-construction failures are excluded. The retained 232,506 Stage-1 structures are deterministically partitioned. The Stage-1 whole-structure convention reverses all moments together when their signed sum is negative:
\begin{equation}
m_i^{(1)}=\kappa_1m_i,\qquad
\kappa_1=\begin{cases}-1,&\sum_i m_i<0,\\1,&\sum_i m_i\geq0.\end{cases}
\label{eq:stage1-gauge}
\end{equation}
Stage~2 instead uses
\begin{equation}
\widetilde m_i=\kappa_{\rm gauge}m_i,\qquad
\kappa_{\rm gauge}=\begin{cases}
-1,&\sum_i m_i<-\delta_{\rm gauge},\\
1,&\sum_i m_i>\delta_{\rm gauge},\\
\kappa_0,&|\sum_i m_i|\leq\delta_{\rm gauge},
\end{cases}
\label{eq:stage2-gauge}
\end{equation}
where $\delta_{\rm gauge}=0.01~\mu_B$ and $\kappa_0$ makes the group with the largest absolute mean moment positive. Both conventions preserve relative signs; near compensation, Stage~2 can leave a small negative sum. Supplementary Note~\suppref{SI-sec:supp-note-1} gives the complete data audit and tie rules.

\subsection{Stage-1 magnetic-moment predictor and combined Stage-2 dataset}\label{sec:stage1-surrogate}

Stage-1 ALIGNN combines periodic bond and bond-angle graphs with standardized elemental descriptors. For minibatch $\mathcal B$, atom count $N_x$ and site prediction $f_{\omega,i}(G_x)$, it minimizes
\begin{equation}
\mathcal L_{\rm node}=\frac{1}{\sum_{x\in\mathcal B}N_x}
\sum_{x\in\mathcal B}\sum_{i=1}^{N_x}
{\rm smooth}L_1\!\left(
\frac{f_{\omega,i}(G_x)-\bar m}{s_m},
\frac{m_i^{(1)}(x)-\bar m}{s_m}\right),
\label{eq:stage1-loss}
\end{equation}
where $\bar m$ and $s_m$ standardize gauge-fixed training-site moments; the SmoothL1 transition is one unit. The checkpoint minimizes validation site-moment MAE.

Its net aggregate has no separate loss:
\begin{equation}
\widehat m_{\rm net}=\frac{1}{N}\left|\sum_i\widehat m_i\right|.
\label{eq:surrogate-aggregate}
\end{equation}
Supplementary Table~\suppref{SI-tabS:stage1} reports fixed validation and test performance; Table~\suppref{SI-tabS:model-settings} lists training settings.

Stage-2 public structures come from NOMAD, OQMD and MP. Within identifier-linked records, DFT-derived labels take precedence through element-resolved periodic-coordinate matching; an element-wise fallback is used only when within-element moments are nearly uniform and cell multiplicities are compatible. Otherwise the fixed ALIGNN predictor supplies gauge-oriented labels.

The combined dataset contains 73,813 records with DFT-derived labels and 3,467,463 with surrogate-predicted labels. Symmetry-equivalent sites define Wyckoff-group targets. Wyckoff groups containing moments both above $0.10~\mu_B$ and below $-0.10~\mu_B$ are excluded from moment-token, sign and value losses; affected structures are excluded from summary loss. The magnitude of the signed site-moment sum per atom defines ten magnetic-condition bins, anchored below $0.5~\mu_B$/atom and at or above $1.8~\mu_B$/atom, with empirical octiles between them.

Training minibatches contain an expected 25\% DFT-labelled fraction, without magnetic-bin reweighting or in-house oversampling. DFT labels receive unit weight; surrogate labels use fixed validation-calibrated reliability weights $(n_{\rm correct}+1/2)/(n+1)$ within each predicted-sign and magnitude bin, where $n_{\rm correct}$ of $n$ Stage-1 validation predictions have the correct sign. Provenance affects sampling and loss weights, not generator inputs or task definitions. Label-transfer, bin-threshold and weighting details are given in Supplementary Notes~\suppref{SI-sec:supp-note-1} and~\suppref{SI-sec:supp-note-2}.

\subsection{Generator, latent banks and GMM sampling}\label{sec:gmmag-generator}

Four branches encode elemental descriptors, space group, Wyckoff group and signed moment (Fig.~\ref{fig:model_architecture}). ALIGNN and bond-graph layers map them to a 128-dimensional diagonal-Gaussian posterior. The ten-bin magnetic condition shifts this posterior and parameterizes a diagonal conditional prior. FiLM conditions decoder hidden representations, while cross-attention reads eight memory vectors projected from $z$.

The token sequence encodes global crystal descriptors followed by Wyckoff identity, species, signed group-mean moment and fractional coordinates for each representative site. Space-group constraints restrict outputs, and padding and inactive fields are excluded from loss.

For active targets $y_i$ with predicted probability $p_i$ and reliability weight $w_i$, define
\begin{align}
\langle q_i\rangle_w&:=\frac{\sum_{i\in\mathcal I}w_i q_i}
{\max(1,\sum_{i\in\mathcal I}w_i)},\label{eq:weighted-average}\\*
h_\delta(e)&:=\begin{cases}e^2/(2\delta),&|e|<\delta,\\ |e|-\delta/2,&|e|\geq\delta,
\end{cases}\label{eq:smooth-l1}
\end{align}
with $\delta=0.1$. The full training objective is
\begin{equation}
\mathcal L_{\rm CVAE}=\mathcal L_{\rm tok}+0.25\mathcal L_{\rm sign}
+0.10\mathcal L_{\rm value}+0.05\mathcal L_{\rm summary}
+10^{-3}\mathcal L_{\rm KL}.
\label{eq:cvae-loss}
\end{equation}
Here $\mathcal L_{\rm tok}=\langle-\ln p_i(y_i)\rangle_w$; the sign, value and summary terms supervise local magnetic-moment signs, expected signed values and $(m_{\rm net},m_{\rm abs},r_{\rm align})$; and $\mathcal L_{\rm KL}$ regularizes the posterior against the conditional prior. The production model was trained with AdamW \cite{loshchilov2019adamw} for 100,000 updates, and its checkpoint was selected by validation loss; matched-budget controls used 560,000 updates. Complete architecture, tokenization, masking, normalization and optimization details are provided in Supplementary Note~\suppref{SI-sec:supp-note-2}.

For generation task $\mathcal S$, training posterior means are standardized coordinate-wise as
\begin{equation}
u_j=\frac{\mu_{q,j}(x,c)-\mu_{{\rm bank},j}}{s_{{\rm std},j}},
\label{eq:latent-standardization}
\end{equation}
where $j=1,\ldots,128$ indexes latent coordinates, and $\mu_{{\rm bank},j}$ and $s_{{\rm std},j}$ are the selected bank's coordinate-wise mean and positive standard deviation. Expectation--maximization fits a regularized diagonal $K$-component GMM to the standardized bank; Eqs.~\eqref{eq:gmm-component}--\eqref{eq:gmm-sampling} sample it with spread scale $\alpha$. Bank structures retain their original encoding conditions, whereas decoding uses the generation-task condition.

The production bank contains 394 Nd-containing structures from condition 8 or 9, excluding Mn and actinides without constraints on $m_{\rm abs}$ or $r_{\rm align}$; the stricter retargeting and matched-control banks use task-specific chemical and magnetic criteria. Most production reconstruction updates used posterior means because training mode was not restored after validation; matched-budget controls restore it.

Targeted samplers share the fixed generator, ALIGNN predictor, formula reference and decoding settings. Each uses five runs of 2,048 draws, retaining all requests in denominators. Figure~\ref{fig:samplers} uses shared principal-component axes and plotting limits; each projected density is normalized to its own maximum, so panels compare location rather than absolute density. Complete bank definitions, screening and sampling protocols are provided in Supplementary Note~\suppref{SI-sec:supp-retargeting}.

\subsection{Finite-budget model and validation design}\label{sec:compute-to-yield-methods}

The novelty reference contains 804,461 canonical reduced formulas from all generator splits. Qualifying reference-unseen draws require valid decoding and geometry, Nd without Mn or actinides, an ALIGNN-predicted aggregate of at least $1.8~\mu_B$/atom and a reference-unseen formula; repetitions still contribute to $A_n$. Calibration uses 45 $(K,\alpha,\tau)$ settings with 50,000 draws each. Latent random-number streams are shared across $\tau$ at fixed $(K,\alpha)$, so those runs are not independent.

Section~\ref{sec:yield-results} defines the joint probability $Q$ of qualification and reference novelty and the distinct-formula accumulation response. The normalized radial distance follows directly from $u=\widehat\mu_k+\alpha L_k\epsilon$, with $L_kL_k^{\mathsf T}=\widehat\Sigma_k$ and $\epsilon\sim\mathcal N(0,I_d)$:
\begin{equation}
\rho_{\rm rms}=\left\{\mathbb E\!\left[
\frac{(u-\widehat\mu_k)^{\mathsf T}\widehat\Sigma_k^{-1}(u-\widehat\mu_k)}{d}
\right]\right\}^{1/2}=\alpha.
\label{eq:rho-alpha}
\end{equation}
The relation $\mathbb E[A_n]=nQ$ assumes a constant joint probability $Q$ throughout a run. We fit Eq.~\eqref{eq:qualifying-response} by penalized product-binomial composite likelihood, fit the accumulation response by robust least squares at observed $A_n$, and combine them through Eq.~\eqref{eq:composed-yield}. Five-fold cross-validation holds out complete sampling settings; parameter diagnostics hold out complete $(\alpha,\tau)$ combinations across all $K$, and final coefficients use all 45 short runs.

Two ten-million-draw runs provide post-fit checks: $K=64$ lies within the calibrated component counts and $K=16$ outside them. Neither is used for fitting, and no prediction extends beyond the observed ten-million-draw boundary. The plotted bands describe block-order and fitted-grid sensitivity rather than confidence intervals or independent-seed variation. The complete calibration grid, fitting diagnostics, dependence structure, numerical approximations and sensitivity analyses are provided in Supplementary Note~\suppref{SI-sec:supp-note-3}.

\subsection{Controlled training and generation under matched protocols}\label{sec:aligned-control-methods}

The magnetic-information comparison trains six configurations from scratch with three seeds to 560,000 updates. Signed-CE\&Aux uses Eq.~\eqref{eq:cvae-loss}; Signed-CE removes its sign, value and summary terms. EncAbs-CE, HistAbs-CE and DualAbs-CE additionally replace signed moments by absolute values only in their designated inputs while retaining signed targets and outputs. These five share partitions, data streams, initialization and optimization within each seed. Global-GReg predicts one continuous nonnegative global moment per atom without local magnetic-moment inputs or outputs, making it a whole-model comparison. The production model is separate.

All 18 models encode the same task-specific bank structures, selected by chemistry and the stored global moment rather than local signs or alignment. Each fits its own standardized $K=64$ diagonal GMM. For every model and task, three shared generation seeds each request 100,000 structures at $(\alpha,\tau)=(1.0,1.2)$ and top-$k=16$. Identical geometry and chemistry checks precede frozen-ALIGNN screening, using $m_{\rm net}\geq1.8~\mu_B$/atom for the Nd and Fe--Co tasks and $m_{\rm net}\leq0.1~\mu_B$/atom for Mn. The predictor receives only the generated crystal, so screening is independent of model-reported moments. Invalid requests remain in denominators.

Decoder-depth and conditioning-path comparisons use the same partitions, 560,000-update budget, three training seeds and final checkpoints. Depth varies over two, four and eight layers at fixed width; path controls retain four layers. Each trained model receives 16,384 attempts per task with one shared generation seed. The four-layer Full baseline reuses the corresponding Signed-CE\&Aux attempts. These controls match updates and generation requests rather than computational cost; parameter counts and removed modules are specified in Supplementary Notes~\suppref{SI-sec:supp-decoder-capacity}--\suppref{SI-sec:supp-decoder-paths}.

Exact magnetic-moment sign agreement uses 4,096 preselected requests per site-output run, independently of screening, novelty and scores. Generated moments $g_i$ are matched to frozen ALIGNN predictions $r_i$ on the same atoms; one whole-structure sign orientation $\eta\in\{-1,+1\}$ minimizes their moment MAE. Exact agreement requires
\begin{equation}
s_\delta(\eta g_i)=s_\delta(r_i)\quad\text{for every }i\text{ with }|r_i|>\delta.
\label{eq:exact-sign-agreement}
\end{equation}
where $s_\delta(x)=-1,0,+1$ for $x<-\delta$, $|x|\leq\delta$ and $x>\delta$, respectively, and $\delta=0.1~\mu_B$. No site-wise flips are allowed. The primary rate includes only geometry- and chemistry-eligible structures whose ALIGNN reference contains both nonzero signs. Reference-zero sites are excluded; a generated zero at a reference-nonzero site is an error. Each eligible occurrence is weighted equally within a run; generation seeds are averaged within training seed and task, the three tasks are equally weighted, and means and sample standard deviations are calculated across training seeds. Architecture comparisons use one generation seed. These metrics compare generated moments with ALIGNN predictions, not with a DFT magnetic ground state. Complete bank, generation, matching and hierarchical averaging rules are provided in Supplementary Note~\suppref{SI-sec:supp-aligned-controls}.
\relax

\subsection{eSEN and DFT evaluation, formula novelty and atomic site occupancy}\label{sec:validation-methods}

One structure with the highest ALIGNN-predicted net aggregate per qualifying reference-unseen formula enters atom-and-cell relaxation with eSEN-30M-OAM \cite{fu_esen}, pretrained on Open Materials 2024 (OMat24) and fine-tuned on Materials Project Trajectory (MPtrj) and subsampled Alexandria (sAlexandria) data \cite{barroso_omat24,deng2023chgnet,schmidt2024alexandria}. DFT selection requires converged relaxation, $E_{\rm form}^{\rm eSEN}<0$ and $M_{\rm pred}/N\geq1.8~\mu_B$/atom, where $M_{\rm pred}=|\sum_i\widehat m_i|$; candidate and reference energies use the same eSEN checkpoint.

Candidate DFT calculations use VASP with spin-polarized PBE \cite{perdew1996pbe}, projector augmented-wave (PAW) datasets \cite{blochl1994paw,kresse1999paw}, a 520~eV cutoff and atom-and-cell relaxation to forces below 0.02~eV~\AA$^{-1}$. One collinear initialization is used without spin--orbit coupling or competing-state enumeration.

For $N=\sum_i n_i$,
\begin{equation}
E_{\rm form}=\frac{E_{\rm cell}-\sum_i n_i\varepsilon_i^{\rm elem}}{N}.
\label{eq:formation-energy}
\end{equation}
Elemental references match PAW and GGA or Dudarev GGA+$U$ conventions \cite{dudarev1998lsdau}. New evaluations preferentially use completed static elemental references; historical results retain their archived matched-reference protocol.

The eSEN convex hull uses DFT-relaxed candidates and one eSEN checkpoint for all candidates and competing phases. For candidate $i$,
\begin{equation}
E_{{\rm hull},i}^{\rm eSEN}=e_i^{\rm eSEN}-\min_{\{\lambda_j\}}\sum_j\lambda_j e_j^{\rm eSEN},
\label{eq:esen-hull}
\end{equation}
where $e_j^{\rm eSEN}$ and $\mathbf{x}_j$ are the per-atom eSEN energy and atomic-fraction vector, subject to $\lambda_j\geq0$, $\sum_j\lambda_j=1$ and $\sum_j\lambda_j\mathbf{x}_j=\mathbf{x}_i$. The catalogue uses relaxed eSEN energies rather than database energies.

Prescribed rare-earth contributions supplement the VASP valence-electron cell spin to estimate collinear saturation magnetization under saturated sublattice alignment. Corrections apply only to recorded fixed-valence PAW datasets omitting the $4f$ shell:
\begin{align}
m_{\rm cell}^{\rm corr}&=m_{\rm VASP}^{\rm valence}+\sum_RN_R\Delta m(P_R),\label{eq:corrected-cell-moment}\\*
\mu_0M_s^{\rm col,est}[\mathrm T]&=11.654\,
\frac{|m_{\rm cell}^{\rm corr}|[\mu_B]}{V_{\rm cell}[\text{\AA}^3]}.\label{eq:collinear-estimate}
\end{align}
$P_R$ identifies the rare-earth PAW dataset and $N_R$ its atom count. Corrections use trivalent Hund-rule projections with prescribed sublattice signs; explicit-valence $4f$ datasets receive none. Potential-specific values are in Supplementary Table~\suppref{SI-tabS:open-core}. This estimate does not determine a magnetic ground state.

Figure~\ref{fig:prospective}b compares this DFT estimate with $\mu_0M_{s,{\rm pred}}^{\rm col,est}=11.654\,|M_{\rm pred}+\sum_RN_R\Delta m(P_R)|/V_{\rm eSEN}$ at the relaxed eSEN geometry.

Reference-unseen status is fixed before DFT inspection and denotes formula novelty only. Fixed held-out and compiled-candidate metrics are deterministic summaries; no null-hypothesis tests are used.

Atomic site-occupancy calculations (Fig.~\ref{fig:abics-disorder}f,g) use abICS on four fixed-composition Nd--Fe--Co parent topologies \cite{liu2024ndfeco,bouchaala2018ndfeco}. Fe/Co occupations vary within the transition-metal sublattice; Nd/Y or Nd/Sm also redistribute within the rare-earth sublattice where present. An on-lattice aenet model guides replica-exchange sampling from 300 to 1800~K, with selected configurations returned to fixed-cell PBE calculations \cite{kasamatsu2022onlattice}. Each system supplies 110 DFT-evaluated configurations across initial sampling and three active-learning rounds. For these site-occupancy calculations, reported energies are final VASP electronic free energies at the chosen smearing, referenced to each system's lowest-energy DFT configuration; only Nd$_2$Fe$_{14}$Co$_3$ has an ordered-parent reference under identical settings. These calculations probe sensitivity to sublattice occupations rather than equilibrium disorder. Supplementary Note~\suppref{SI-sec:supp-note-4} provides complete protocols for relaxation, DFT, elemental references, convex-hull construction and sampling, together with sensitivity audits.
\relax

\section{Acknowledgements}

This work was partly supported by JSPS KAKENHI Grant Number JP24K01144, MEXT-DXMag Grant Number JPMXP1122715503 and JST SPRING Grant No. JPMJSP2180. The funders provided financial support for computational resources. The calculations were partly carried out using supercomputers at the Institute for Solid State Physics of the University of Tokyo and the TSUBAME4.0 supercomputer at Institute of Science Tokyo.

\section{Data availability}

Public datasets are available from the cited repositories. The in-house Nd dataset and pretrained weights are available from the authors upon request.

\section{Code availability}

Code is available from the authors upon request.

\section{Author contributions}

W.-C.S. conceived the study, developed the workflow, performed the computational experiments, analysed the data and drafted the manuscript. C.S. contributed to data interpretation, read the manuscript and provided suggestions. M.T. and T.T. performed part of the DFT calculations and contributed part of the dataset. Y.G. contributed to study design, interpretation and supervision and provided computational resources. All authors reviewed the manuscript.

\section{Competing interests}

All authors declare no financial or non-financial competing interests.

\begingroup
\addtolength{\bibsep}{-2.5pt}
\putbib[references]
\endgroup
\end{bibunit}

\clearpage
\makeatletter
\let\title\GMMagTitle
\let\author\GMMagAuthor
\let\maketitle\GMMagMakeTitle
\let\@maketitle\GMMagInternalMakeTitle
\let\caption\GMMagSICaption
\let\@caption\GMMagSIInternalCaption
\let\@xfloat\GMMagSIXFloat
\let\@floatboxreset\GMMagSIFloatReset
\global\aucount=0
\global\corraucount=0
\global\addcount=0
\global\emailcnt=0
\global\punctcount=0
\gdef\artauthors{}
\gdef\auaddress{}
\gdef\@auaddress{}
\gdef\authorsep{}
\gdef\sep{}
\gdef\@corref{}
\gdef\@abstract{}
\gdef\@keywords{}
\gdef\authemail{}
\gdef\corrauthemail{}
\@fptop=\GMMagFloatTop
\@fpbot=\GMMagFloatBottom
\makeatother
\setcounter{page}{1}
\setcounter{section}{0}
\setcounter{subsection}{0}
\setcounter{subsubsection}{0}
\setcounter{figure}{0}
\setcounter{table}{0}
\setcounter{equation}{0}
\setcounter{footnote}{0}
\setcounter{affn}{0}
\setlength{\LTcapwidth}{\textwidth}
\hypersetup{linkcolor=black,filecolor=black}
\normalsize

\makeatletter
\def\@extra@binfo{.SI}
\def\@extra@b@citeb{.SI}
\makeatother
\begin{bibunit}[sn-nature]

\title{Supplementary Information for Magnetic crystal generation with site-resolved magnetic moments via Gaussian-mixture retargeting and finite-budget yield prediction}

\author[1]{\fnm{Wen-Chuang} \sur{Shang}}
\author[1]{\fnm{Chenyu} \sur{Shen}}
\author[1]{\fnm{Mikito} \sur{Tsujimoto}}
\author[1]{\fnm{Tomonori} \sur{Tanaka}}
\author*[1]{\fnm{Yoshihiro} \sur{Gohda}}

\affil[1]{\orgdiv{Department of Materials Science and Engineering}, \orgname{Institute of Science Tokyo}, \orgaddress{\city{Yokohama} \postcode{226-8501}, \country{Japan}}\\
\textcolor{blue}{\href{mailto:shang.w.beed@m.isct.ac.jp}{shang.w.beed@m.isct.ac.jp}}; \textcolor{blue}{\textsuperscript{*}\href{mailto:gohda@mct.isct.ac.jp}{gohda@mct.isct.ac.jp}}}

\maketitle

\renewcommand{\thefigure}{S\arabic{figure}}
\renewcommand{\thetable}{S\arabic{table}}
\renewcommand{\theequation}{S\arabic{equation}}

\section*{Scope and notation}

These six Supplementary Notes define and evaluate GMMag: Stage~1 predicts site-resolved magnetic moments, and Stage~2 generates crystal structures and magnetic moments. Magnetic screening quantities for generated structures use the frozen ALIGNN predictor of site-resolved magnetic moments unless identified as DFT results. Generated moment tokens are identified separately. Retargeting, large-scale generation and physical evaluation use the production checkpoint; the architecture and magnetic-information controls are trained separately at a common update budget in Notes~\ref{SI-sec:supp-note-2} and~\ref{SI-sec:supp-aligned-controls}; their outcomes are not pooled.

\section{Data and magnetic labels}\label{SI-sec:supp-note-1}

\subsection{Composition and label provenance}

DFT-derived local magnetic-moment labels combine AFLOW \cite{curtarolo2012aflow}, OQMD \cite{kirklin2015oqmd} and in-house VASP calculations \cite{kresse1996vasp}; generator structures additionally come from Materials Project \cite{jain2013materialsproject} and the Novel Materials Discovery repository (NOMAD) \cite{draxl2018nomad}. Table~\ref{SI-tabS:data-summary} distinguishes structural origin, label provenance and identifier overlap. Retained public labels are collinear and spin-polarized, with no spin--orbit coupling in the available metadata; orbital moments are not separately labelled. Public atomic projections lack a common recorded partition or projection radius, so their sums are learning and screening aggregates rather than complete-cell magnetic moments. The combined dataset with DFT-derived and predicted magnetic labels includes 174 eligible in-house Nd-containing structures; reported performance concerns the integrated workflow.

\begin{table}[!htbp]
\centering\small
\caption{Corpus composition and cross-stage overlap. a, Counts by source, label provenance and partition. b, DFT-derived/surrogate-predicted Stage-2 label counts matched to Stage-1 records by identifier. Identifier overlap does not establish structural or formula-level independence.}
\label{SI-tabS:data-summary}\label{SI-tabS:cross-stage}
\begin{tabular}{llcrrr}
\toprule
\multicolumn{6}{l}{\textbf{a, Source and label provenance}}\\
Stage & Source & Labels & Training & Validation & Test\\\midrule
1 & AFLOW & DFT & 171,138 & 21,392 & 21,393 \\
1 & OQMD & DFT & 14,660 & 1,832 & 1,833 \\
1 & In-house & DFT & 202 & 27 & 29 \\
2 & Materials Project & Predicted & 106,046 & 11,670 & 625 \\
2 & NOMAD & Predicted & 2,181,922 & 242,399 & 12,230 \\
2 & NOMAD & DFT & 65,834 & 7,460 & 345 \\
2 & OQMD & Predicted & 817,219 & 90,836 & 4,516 \\
2 & In-house & DFT & 135 & 18 & 21 \\\bottomrule
\relax
\end{tabular}
\medskip
\begin{tabular}{lrrr}
\toprule
\multicolumn{4}{l}{\textbf{b, Identifier overlap}}\\
Stage-1 partition & Stage-2 training & Stage-2 validation & Stage-2 test\\\midrule
Training & 52,873 / 19,153 & 5,987 / 2,143 & 290 / 105 \\
Validation & 6,515 / 2,348 & 764 / 262 & 29 / 19 \\
Test & 6,581 / 2,303 & 727 / 261 & 47 / 12 \\\bottomrule
\relax
\end{tabular}
\end{table}

Stage~1 orients the moment vector to a nonnegative site sum; Stage~2 applies a Wyckoff-group tie-break within $0.01~\mu_B$ of compensation (main-text Eqs.~\MainEq{eq:stage1-gauge} and \MainEq{eq:stage2-gauge}). Magnitude differences within $10^{-8}~\mu_B$ are ties, resolved by species, Wyckoff identity and wrapped fractional coordinates; $\kappa_0=1$ if the largest absolute group mean is at most $10^{-12}~\mu_B$. Relative signs are preserved: 61,137 Stage-1 and 2,453,025 Stage-2 structures retain a site below $-10^{-6}~\mu_B$. The 144,795 negative Stage-2 totals all lie in $[-0.01,0)~\mu_B$.

Geometry and moment arrays must have consistent counts and ordering; in-house geometry--moment pairs come from the same completed calculation. Non-finite arrays, inconsistent lengths and graph-construction failures are excluded. Transfer of DFT-derived labels uses 74,349 identifier-linked public records. Element-resolved periodic-coordinate matching permits a wrapped-coordinate difference of 0.002, first in the stored cell and then after symmetry standardization at 0.01~\AA. Labels are averaged only when their spread is at most $0.005~\mu_B$. An element-wise fallback permits means at the same tolerance when elemental counts share one positive integer cell-multiplicity ratio; it does not assert structural equivalence. Assignments comprise 5,154 direct-cell, 57,153 standardized-cell and 11,332 element-wise matches. The 710 failures and 25,896 identifier-linked records outside this pool retain surrogate-predicted labels. Cross-source symmetry equivalence and formula overlap have not been exhaustively tested.

\subsection{Prediction, reconstruction and magnetic conditions}

The Stage-1 ALIGNN surrogate \cite{choudhary2021alignn} is evaluated on fixed validation and test sets, with training settings in Table~\ref{SI-tabS:model-settings}. The signed sum denotes the oriented sum of site moments, whereas the absolute sum adds their magnitudes. Both are unnormalized sums in $\mu_B$, not moments per atom or complete-cell DFT moments. Table~\ref{SI-tabS:stage1} also reports the production generator's held-out reconstruction, with its protocol in Note~\ref{SI-sec:supp-generator-heldout}.

\begin{table}[!htbp]
\centering\small
\caption{Held-out prediction and reconstruction. a, Stage-1 errors and coefficients of determination for local magnetic moments and their sums on fixed splits; all MAEs are in $\mu_B$. b, Posterior-mean, teacher-forced reconstruction of 4,096 test structures from the combined dataset by the production generator. Validity, composition match and sign accuracy are fractions. Group metrics refer to Wyckoff-site groups; sign accuracy uses groups with nonzero target moments ($|m|>0.10~\mu_B$). These are deterministic diagnostics, separate from free-generation evaluation.}
\label{SI-tabS:stage1}\label{SI-tabS:models}
\setlength{\tabcolsep}{3pt}
\begin{tabular}{lrrrrrr}
\toprule
\multicolumn{7}{l}{\textbf{a, Magnetic surrogate}}\\
Split & \shortstack{Site\\MAE} & \shortstack{Site\\$R^2$} & \shortstack{Signed sum\\MAE} & \shortstack{Signed sum\\$R^2$} & \shortstack{Absolute sum\\MAE} & \shortstack{Absolute sum\\$R^2$}\\\midrule
Validation & 0.05 & 0.889 & 0.16 & 0.980 & 0.16 & 0.983 \\
Test & 0.05 & 0.899 & 0.16 & 0.978 & 0.17 & 0.979 \\
\bottomrule
\relax
\end{tabular}
\medskip
\begin{tabular}{lrrrr}
\toprule
\multicolumn{5}{l}{\textbf{b, Production-model reconstruction}}\\
Generator & \shortstack{Structure\\valid} & \shortstack{Composition\\match} & \shortstack{Group MAE\\($\mu_B$)} & \shortstack{Nonzero-group\\sign accuracy}\\\midrule
Main ALIGNN encoder & 0.958 & 0.983 & 0.01 & 0.959 \\
\bottomrule
\relax
\end{tabular}
\end{table}

The ten magnetic conditions are anchored at 0.5 and 1.8~$\mu_B$/atom. Seven empirical octile boundaries lie between them, approximately $(0.57,\allowbreak 0.65,\allowbreak 0.73,\allowbreak 0.82,\allowbreak 0.91,\allowbreak 1.01,\allowbreak 1.23)~\mu_B$/atom. The octiles come from the original actinide-free training distribution with surrogate-predicted labels and are retained for the expanded corpus; condition assignment uses the exact thresholds in the Source Data.

The combined dataset contains 73,813 records with DFT-derived labels and 3,467,463 with surrogate-predicted labels. Training minibatches have an expected 25\% DFT-labelled fraction without magnetic-bin reweighting or in-house oversampling; validation and test distributions remain fixed. Its 116 Wyckoff groups containing both nonzero sign classes in 113 structures are excluded from moment-token, sign and value losses; affected structures are excluded from summary loss, as defined in Methods.
\FloatBarrier
\relax
\section{Generator specification and reconstruction}\label{SI-sec:supp-note-2}

The CVAE with site-resolved magnetic moments \cite{kingma2014vae,sohn2015cvae} uses a periodic directed bond graph $G$ and bond-angle line graph $L(G)$. At most 12 neighbours lie within 8~\AA; each directed bond is a line-graph node, and consecutive bonds define an angle. The expressions below expand the main-text objective and specify its masks and numerical implementation.

\subsection{Input branches and graph encoder}

For atom $i$, the basic vector $b_i$ contains standardized atomic number, periodic-table group and period, electronegativity, atomic radius, $s/p/d/f$ valence counts, first ionization energy and electron affinity, also used by the Stage-1 predictor. Missing generator elemental values are mean-imputed before standardization. Using Stage-2 moment-normalization statistics $\bar m_{\rm gen},s_{m,{\rm gen}}$ and gauge-fixed $\tilde m_i$, the four input branches are:
\begin{align}
h_i^{\rm basic}&={\rm MLP}_{11\rightarrow64\rightarrow64}(b_i),\\
h_i^{\rm SG}&={\rm MLP}_{64\rightarrow64\rightarrow64}(e_{\rm SG}),\\
h_i^{\rm Wy}&={\rm MLP}_{64\rightarrow64\rightarrow64}(e_{\rm Wy}),\\
h_i^{m}&={\rm MLP}_{1\rightarrow64\rightarrow64}\!\left[(\tilde m_i-\bar m_{\rm gen})/s_{m,{\rm gen}}\right].
\end{align}
Their direct concatenation gives $h_i^{(0)}=[h_i^{\rm basic}\Vert h_i^{\rm SG}\Vert h_i^{\rm Wy}\Vert h_i^m]\in\mathbb R^{256}$. Global-GReg sets $h_i^m$ to zero; its other changes are specified in Note~\ref{SI-sec:supp-aligned-controls}. Bond distances use 80 radial basis functions on $[0,8]$~\AA{} and angle cosines use 40 basis functions on $[-1,1]$; each expansion is mapped to 256 dimensions.

The edge-gated update for a directed edge $i\rightarrow j$ is
\begin{align}
u_{ij}&=W_g^{\rm src}h_i+W_g^{\rm dst}h_j+W_g^e e_{ij},\\
g_{ij}&=\operatorname{sigmoid}(u_{ij}),\\
r_j&=W_n^{\rm self}h_j+
\frac{\sum_{i\rightarrow j}g_{ij}\odot W_n^{\rm nbr}h_i}
{\sum_{i\rightarrow j}g_{ij}+10^{-6}\mathbf 1},\\
h_j'&=h_j+\operatorname{SiLU}[{\rm BN}(r_j)],\\
e_{ij}'&=e_{ij}+\operatorname{SiLU}[{\rm BN}(u_{ij})].
\end{align}
The quotient in the $r_j$ update is evaluated elementwise.
Four ALIGNN layers \cite{choudhary2021alignn} alternately update atom--bond states on $G$ and bond--angle states on $L(G)$; four further edge-gated convolutions operate on $G$. Mean pooling produces $g_x=N_x^{-1}\sum_i h_i^{(8)}$. The CGCNN \cite{xie2018cgcnn} and DimeNet++ \cite{gasteiger2020dimenetpp} comparisons use eight distance-based or eight triplet-interaction blocks, respectively, retaining the same 256-dimensional pooled representation and latent and decoder interfaces.

\subsection{Conditional latent distribution}

Linear maps of $g_x$ produce base posterior parameters $\mu_x^0,\ell_x^0\in\mathbb R^{128}$. For magnetic-condition bin $c\in\{0,\ldots,9\}$, a separate embedding and MLP produce posterior offsets:
\begin{align}
\mu_q(x,c)&=\mu_x^0+\Delta\mu_c,\\
\ell_q(x,c)&=\ell_x^0+\Delta\ell_c,\\
q_\phi(z|x,c)&=\mathcal N[\mu_q,\operatorname{diag}(\exp\ell_q)],\\
z&=\mu_q+\exp(\ell_q/2)\odot\epsilon,
\end{align}
where $\epsilon\sim\mathcal N(0,I)$ is a standard normal noise vector. A distinct condition-bin embedding and three-layer MLP parameterize the conditional prior $p_\psi(z|c)=\mathcal N[\mu_p(c),\operatorname{diag}(\exp\ell_p(c))]$, with $\ell_p\in[-4,0]^{128}$. No auxiliary magnetic-condition classifier is attached to $z$.

\subsection{Signed tokenization, decoder and allowed outputs}\label{SI-sec:supp-decoder-operations}

The fixed causal target has length 202:
\begin{equation}
\begin{aligned}
y={}&\bigl[\mathrm{SG},n_{\rm elem},N_{\rm full},a_{\rm lat},b_{\rm lat},c_{\rm lat},\alpha_{\rm lat},\beta_{\rm lat},\gamma_{\rm lat},N_{\rm rep},\\
&\qquad\{w_r,Z_r,m_r,f_{r1},f_{r2},f_{r3}\}_{r=1}^{32}\bigr].
\end{aligned}
\end{equation}
The ten global fields encode space group, element count, full-cell atom count, lattice lengths and angles, and representative-group count. Each of up to 32 site blocks gives Wyckoff identity $w_r$, species $Z_r$, signed group-mean moment $m_r$ and fractional coordinates $f_{r1},f_{r2},f_{r3}$. Padding and sign-conflict exclusions follow the main-text masks, and space-group constraints retain only independent lattice quantities and free coordinate axes. Quantizers and losses use numerical moments, summary moments, lengths and angles in $\mu_B$, $\mu_B$/atom, \AA{} and degrees, respectively; $r_{\rm align}$ is dimensionless. The categorical maps are
\begin{align}
Q_f(f)&=\operatorname{clip}(\operatorname{round}[100(f\bmod1)],0,99),\\
Q_m(m)&=\operatorname{clip}(\operatorname{round}[100(m+7)],0,1400),\\
Q_l(l)&=\operatorname{clip}(\operatorname{round}[l/0.01],0,4000),\\
Q_\vartheta(\vartheta)&=\operatorname{clip}(\operatorname{round}[(\vartheta-30)/0.5],0,240).
\end{align}
Moments span $[-7,7]~\mu_B$ at $0.01~\mu_B$ resolution; lattice lengths and angles use 0.01~\AA{} and $0.5^\circ$. Moments below $0.01~\mu_B$ in magnitude map to zero before quantization. Coordinate rounding to 100 is clipped to token 99 rather than wrapped again to zero. Teacher forcing shifts targets right and combines token, slot-type and position embeddings, followed by layer normalization.

The pre-normalized transformer \cite{vaswani2017attention} receives latent information through cross-attention and the magnetic-condition bin through feature-wise linear modulation (FiLM) \cite{perez2018film}. A two-layer MLP with GELU activation projects the CVAE latent vector into memory:
\begin{equation}
M_z=\operatorname{reshape}_{8\times256}
\!\left[\operatorname{MLP}_{128\rightarrow1024\rightarrow2048}(z)\right].
\label{SI-eq:supp-latent-memory}
\end{equation}
The rows of $M_z$ are eight learned memory vectors, not individual atomic sites. Let $H^{(l-1)}$ contain the token hidden representations entering layer $l$. Masked self-attention and its residual connection give $S^{(l)}=H^{(l-1)}+\operatorname{Dropout}\{\operatorname{SelfAttn}[\operatorname{LN}(H^{(l-1)})]\}$; the causal mask prevents access to future tokens. For cross-attention head $h$, the queries $Q_h^{(l)}$ are learned affine projections of $\operatorname{LN}(S^{(l)})$, while keys $K_h^{(l)}$ and values $V_h^{(l)}$ are separate learned affine projections of $M_z$. Each head computes
\begin{equation}
A_h^{(l)}=\operatorname{softmax}\!\left[
\frac{Q_h^{(l)}(K_h^{(l)})^{\mathsf T}}{\sqrt{d_h}}
\right]V_h^{(l)},
\label{SI-eq:supp-cross-attention}
\end{equation}
where $d_h=32$, $h=1,\ldots,8$, and softmax normalizes over the eight memory vectors. Combining heads and adding the feed-forward sublayer gives
\begin{align}
C^{(l)}&=S^{(l)}+\operatorname{Dropout}\!\left[
\operatorname{Concat}(A_1^{(l)},\ldots,A_8^{(l)})W_O^{(l)}+b_O^{(l)}
\right],\label{SI-eq:supp-cross-residual}\\
\widetilde H^{(l)}&=C^{(l)}+\operatorname{Dropout}\!\left[
\operatorname{FFN}^{(l)}(\operatorname{LN}(C^{(l)}))
\right].\label{SI-eq:supp-decoder-ffn}
\end{align}
Here LN is layer normalization and FFN is the $256\rightarrow1024\rightarrow256$ feed-forward network with GELU activation. The attention and feed-forward sublayers also use dropout during training; it is omitted from the head expression above for clarity.

For condition bin $c\in\{0,\ldots,9\}$, its learned embedding $e_c\in\mathbb R^{256}$ enters a separate MLP in each layer to produce raw scale and shift vectors $(\widetilde g_c^{(l)},\widetilde b_c^{(l)})$. FiLM acts after the entire decoder layer:
\begin{align}
g_c^{(l)}&=\mathbf1+0.1\tanh\widetilde g_c^{(l)},\label{SI-eq:supp-film-scale}\\
b_c^{(l)}&=0.1\tanh\widetilde b_c^{(l)},\label{SI-eq:supp-film-shift}\\
H_t^{(l)}&=g_c^{(l)}\odot\widetilde H_t^{(l)}+b_c^{(l)}.\label{SI-eq:supp-film}
\end{align}
The same 256 scale and shift values apply to every token position $t$ within a layer. They depend on $c$, not on $z$; FiLM modifies the token hidden representations, leaving the sampled latent vector and its projected memory unchanged. The mainline has four such layers (Table~\ref{SI-tabS:model-settings}).

Field-specific heads predict space group (231 classes), unique-element count (17), full-cell count (1249), representative-group count (33), Wyckoff identity (28), species (119), signed moment (1,401), fractional coordinate (100), lattice length (4001) and angle (241). Wyckoff logits are masked using the generated space group, remaining multiplicity, remaining representative-group budget and exact reachability of the requested full-cell count.

\subsection{Objective and optimization}

Let $\mathcal A_0(x)$ contain active non-moment positions, $\mathcal M(x)$ the active moment groups without conflicting signs, and $t_r$ the moment-token position of group $r$. DFT-derived labels have reliability weight $\omega_{xr}=1$. For surrogate-predicted labels, each predicted-sign and magnitude bin uses $\omega_{xr}=(n_{\rm correct}+1/2)/(n+1)$, the Jeffreys posterior mean from Stage-1 validation data. Magnitude-bin edges are $[0,0.05,0.10,0.25,0.50,1.0,1.5,2.5,4.0,8.0]~\mu_B$; bins with fewer than 25 observations use the corresponding sign-class mean, and weights are clipped to $[0.25,0.995]$. The reliability-weighted token cross-entropy expands to
\begin{align}
\mathcal N_{\rm tok}
&=-\sum_{x\in\mathcal B}\sum_{t\in\mathcal A_0(x)}
\ln p_{\theta,\operatorname{head}(t)}(y_t|y_{<t},z,c)\nonumber\\
&\quad-\sum_{x\in\mathcal B}\sum_{r\in\mathcal M(x)}\omega_{xr}
\ln p_{\theta,m}(Q_m(m_{xr})|y_{<t_r},z,c),\\
\mathcal W_{\rm tok}
&=\sum_{x\in\mathcal B}|\mathcal A_0(x)|
+\sum_{x\in\mathcal B}\sum_{r\in\mathcal M(x)}\omega_{xr},\\
\mathcal L_{\rm tok}&=\mathcal N_{\rm tok}/\max\{1,\mathcal W_{\rm tok}\}.
\end{align}
Here $\operatorname{head}(t)$ selects the field-specific output head at token position $t$, and $t_r$ is the moment-token position of group $r$.

For the auxiliary terms, let $v_j=(-7+0.01j)\,\mu_B$, $j=0,\ldots,1400$, and let $p_{xrj}$ be the softmax probability of moment class $v_j$. The expected signed moment is
\begin{equation}
\widehat m_{xr}=\sum_{j=0}^{1400}p_{xrj}v_j.
\end{equation}
The sign groups are $C_- =\{j:v_j<-0.10\}$, $C_0=\{j:|v_j|\leq0.10\}$ and $C_+=\{j:v_j>0.10\}$, with $P_{xr,g}=\sum_{j\in C_g}p_{xrj}$ for $g\in\{-,0,+\}$. Their logits are computed by a log-sum-exp over each group. If $y_{xr}^{\rm sign}\in\{-,0,+\}$ is the target class, the sign loss is
\begin{equation}
\mathcal L_{\rm sign}=
\frac{\sum_{x\in\mathcal B}\sum_{r\in\mathcal M(x)}
\omega_{xr}w_{y_{xr}^{\rm sign}}^{\rm cls}[-\ln P_{xr,y_{xr}^{\rm sign}}]}
{\max\!\left\{1,\sum_{x\in\mathcal B}\sum_{r\in\mathcal M(x)}\omega_{xr}\right\}},
\end{equation}
where $(w_-^{\rm cls},w_0^{\rm cls},w_+^{\rm cls})\simeq(2.085,\,0.251,\,0.664)$. These fixed square-root inverse-frequency multipliers are normalized to unit mean, clipped to $[0.25,6]$, then renormalized. Displayed values are rounded; full precision is retained in the Source Data.

Using the smooth-$L_1$ function in main-text Eq.~\MainEq{eq:smooth-l1}, with $\delta_{\rm SL1}=0.10$, the expected-value loss is
\begin{equation}
\mathcal L_{\rm value}=
\frac{\sum_{x\in\mathcal B}\sum_{r\in\mathcal M(x)}
\omega_{xr}h_{\delta_{\rm SL1}}(\widehat m_{xr}-m_{xr})}
{\max\!\left\{1,\sum_{x\in\mathcal B}\sum_{r\in\mathcal M(x)}\omega_{xr}\right\}}.
\end{equation}

Let $\nu_{xr}$ be the Wyckoff-site-group multiplicity and $N_x^{\rm mom}=\max\{1,\sum_{r\in\mathcal M(x)}\nu_{xr}\}$. The predicted summaries of site moments, normalized per atom, and their dimensionless ratio are
\begin{align}
\widehat m_{{\rm net},x}&=(N_x^{\rm mom})^{-1}\left|\sum_{r\in\mathcal M(x)}\nu_{xr}\widehat m_{xr}\right|,\label{SI-eq:supp-predicted-summaries}\\
\widehat m_{{\rm abs},x}&=(N_x^{\rm mom})^{-1}\sum_{r\in\mathcal M(x)}\nu_{xr}|\widehat m_{xr}|,\\
\widehat r_{{\rm align},x}&=\frac{\widehat m_{{\rm net},x}}
{\max(\widehat m_{{\rm abs},x},\varepsilon_m)}.
\end{align}
The floor $\varepsilon_m=10^{-6}~\mu_B$/atom gives $\widehat r_{{\rm align},x}=0$ for an exactly zero predicted vector. The stored target follows main-text Eq.~\MainEq{eq:magnetic-alignment}, with a zero ratio for an exactly zero moment vector.
Let $\mathcal B_{\rm sum}\subseteq\mathcal B$ contain batch structures without a conflicting group, let $\bar\omega_x=|\mathcal M(x)|^{-1}\sum_{r\in\mathcal M(x)}\omega_{xr}$ and define $W_{\rm sum}=\sum_{x\in\mathcal B_{\rm sum}}\bar\omega_x$. The summary loss averages the three component errors before reliability weighting:
\begin{align}
e_x&=\frac13\Bigl[
h_{\delta_{\rm SL1}}(\widehat m_{{\rm net},x}-m_{{\rm net},x})
+h_{\delta_{\rm SL1}}(\widehat m_{{\rm abs},x}-m_{{\rm abs},x})\nonumber\\
&\hspace{16mm}+h_{\delta_{\rm SL1}}(\widehat r_{{\rm align},x}-r_{{\rm align},x})\Bigr],\\
\mathcal L_{\rm summary}&=
\frac{\sum_{x\in\mathcal B_{\rm sum}}\bar\omega_x e_x}
{\max\{1,W_{\rm sum}\}}.\label{SI-eq:supp-summary-loss}
\end{align}

For diagonal posterior and conditional-prior distributions, $\ell_q$ and $\ell_p$ denote log-variances. Clipping bounds are given in Table~\ref{SI-tabS:model-settings}. The per-structure Kullback--Leibler (KL) divergence and its batch mean are
\begin{align}
\mathcal K_x&=\frac12\sum_{j=1}^{128}
[\ell_{p,j}-\ell_{q,j}+\exp(\ell_{q,j}-\ell_{p,j})\nonumber\\
&\hspace{24mm}+(\mu_{q,j}-\mu_{p,j})^2\exp(-\ell_{p,j})-1],\\
\mathcal L_{\rm KL}&=|\mathcal B|^{-1}\sum_{x\in\mathcal B}\mathcal K_x.
\end{align}
The full objective is main-text Eq.~\MainEq{eq:cvae-loss}. Table~\ref{SI-tabS:model-settings} distinguishes the production and matched-budget training schedules; sampling balanced by label provenance and reliability weighting act separately. Signed-CE and the absolute-input controls in Note~\ref{SI-sec:supp-aligned-controls} retain token weighting while removing the three auxiliary coefficients.

\begin{table}[!htbp]
\centering\small
\caption{Model and optimization settings. The Stage-1 predictor and Stage-2 generator are trained separately. Production training selects a checkpoint by validation loss; all six full-training controls use final checkpoints at the common update budget. Generator defaults apply to both training protocols.}
\label{SI-tabS:model-settings}
\renewcommand{\arraystretch}{1.0}
\begin{tabular}{@{}>{\raggedright\arraybackslash}p{0.27\textwidth}>{\raggedright\arraybackslash}p{0.68\textwidth}@{}}
\toprule
Component & Setting\\\midrule
Stage-1 graph and head & At most 12 neighbours within 8~\AA; radial and angular bases; four ALIGNN layers, four bond-graph convolutions and a multilayer-perceptron node head\\
Stage-1 training & 300,000 updates; select lowest predefined validation site-moment MAE\\
Generator decoder & Four layers; eight attention heads; width 256; feed-forward width 1024\\
Latent memory & $128\to1024\to(8\times256)$ projection; eight memory vectors\\
KL evaluation & Posterior log-variance clipped to $[-20,20]$; prior to $[-8,8]$ (the prior parameterization itself restricts log-variance to $[-4,0]$)\\
Generator defaults & AdamW; batch size 32; dropout 0.1; weight decay 0.01\\
Production model & 100,000 updates; learning rates $5\times10^{-5}$ (moment-token embeddings, moment head, shared token embedding) and $10^{-5}$ (other parameters); 2,000 warm-up updates; linear decay to 0.2 of the initial rates\\
Full-training controls & 560,000 updates; learning rate $10^{-4}$; 1,000 warm-up updates; cosine decay to 0.2 of the initial rate\\
\bottomrule
\end{tabular}
\end{table}
\relax
\subsection{Held-out reconstruction}\label{SI-sec:supp-generator-heldout}

Production-model reconstruction (Table~\ref{SI-tabS:models}b) uses the first 4,096 test records from the combined dataset, posterior means, the true preceding tokens (teacher forcing) and field-wise maximum-probability tokens. Group-mean magnetic-moment MAE and sign accuracy exclude padded and sign-conflicting groups. Group sign accuracy uses target moments with $|m|>0.10~\mu_B$ and compares target and predicted signs using the same three sign classes as the sign loss; a predicted zero is an error. No global sign reversal is applied. Batch-level metrics receive equal weight, including the last partial batch; empty sign denominators contribute zero. Structural validity uses a minimum interatomic-separation tolerance of $0.5$~\AA{}. Freely generated structures are evaluated separately with the frozen ALIGNN predictor.
\FloatBarrier

\subsection{Compact encoder comparison}\label{SI-sec:supp-encoder-comparison}

\begin{figure}[!htbp]
\centering
\includegraphics[width=\textwidth]{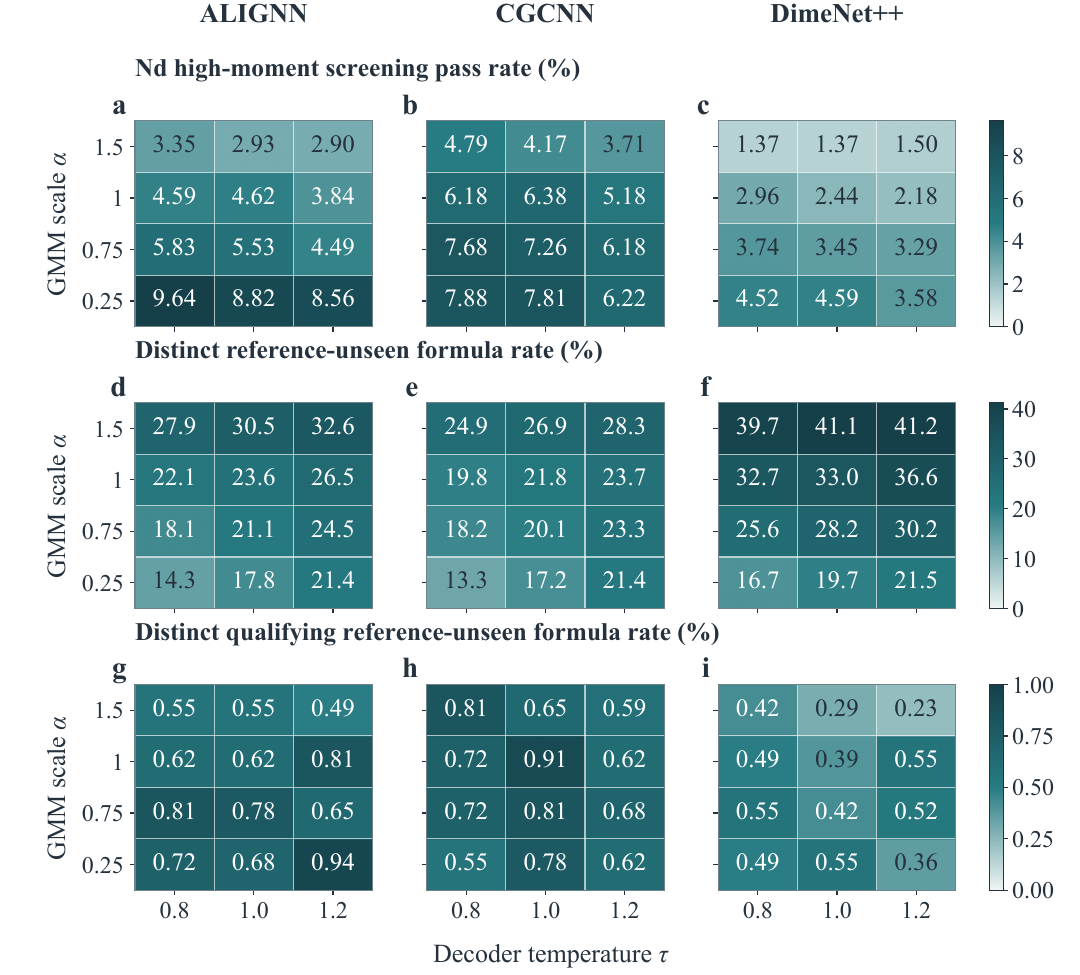}
\caption{Matched generation sweeps for ALIGNN, CGCNN and DimeNet++ CVAEs in the Nd-containing high-moment generation task. Columns correspond to the three encoders. Rows show the screening pass rate before novelty filtering, distinct reference-unseen formula rate without screening, and distinct qualifying reference-unseen formula rate; all denominators are requested draws. Screening requires valid decoding, the 0.5~\AA{} geometry check, Nd presence, Mn/actinide absence and surrogate $m_{\rm net}\geq1.8~\mu_B$/atom, without a net-to-absolute moment ratio requirement. Each model independently encodes the same 394 production-bank records. Decoding uses condition 9 and top-$k=16$; latent sampling uses a $K=64$ GMM. Each cell averages three shared generation seeds with 1,024 requests per seed, at the displayed GMM scale $\alpha$ and decoder temperature $\tau$. Each encoder contributes one training checkpoint.}
\label{SI-figS:encoder-sweep}
\end{figure}

Figure~\ref{SI-figS:encoder-sweep} compares one independently trained checkpoint with signed site-resolved moments per encoder in the Nd-containing high-moment generation task. Narrowing the GMM components by reducing the sampling scale increases the screening pass rate, whereas increasing the scale expands formula-level diversity. DimeNet++ gives the largest distinct reference-unseen formula rate, while ALIGNN and CGCNN give similar, higher yields of distinct qualifying reference-unseen formulas. These encoder-checkpoint comparisons are separate from the magnetic controls trained at a common update budget in Note~\ref{SI-sec:supp-aligned-controls}; shared generation seeds assess sampling variation within each checkpoint.

\FloatBarrier
\subsection{Decoder capacity}\label{SI-sec:supp-decoder-capacity}

The depth comparison uses two, four or eight transformer decoder layers, including one FiLM conditioner per layer, with the same ALIGNN encoder, token representation and training objective. Width, attention-head count, feed-forward width and latent dimension remain 256, 8, 1,024 and 128, respectively. Each configuration is trained from random initialization in three training instances to 560,000 updates with batches of 32. The four-layer model reuses the Signed-CE\&Aux controls in Note~\ref{SI-sec:supp-aligned-controls}, not the historical production checkpoint. The comparison matches training updates and generation requests rather than computation time.

Table~\ref{SI-tabS:decoder-capacity} evaluates the same 17,737 held-out records using posterior means, true preceding tokens (teacher forcing) and field-wise maximum-probability predictions. Composition matching requires identical element counts after Wyckoff multiplicities are included. The lattice error concerns length $a$ only; Wyckoff parameter error is the periodic fractional-coordinate MAE over target-active parameters. Test metrics average 555 batches equally, including the final nine-record batch, except that Wyckoff MAE pools active-coordinate errors. These diagnostics measure conditional token and parameter prediction, separately from free generation in main-text Fig.~\ref{fig:decoder-depth}.

\begin{table}[!htbp]
\centering\small
\caption{Decoder capacity and held-out prediction. a, Trainable parameter counts in millions. Transformer blocks include self-attention, cross-attention, feed-forward networks, layer normalization and per-layer FiLM; embeddings, memory projection, output heads and the encoder are excluded from this column and included in the total model count. b, Mean $\pm$ sample standard deviation across three training instances. Composition matching is a percentage; $a$ MAE is in \AA{} and Wyckoff parameter MAE is dimensionless.}
\label{SI-tabS:decoder-capacity}
\setlength{\tabcolsep}{5pt}
\begin{tabular}{lrr}
\toprule
\multicolumn{3}{l}{\textbf{a, Parameter counts}}\\
Layers & Transformer blocks & Total model\\\midrule
2 & 2.30 & 21.31\\
4 & 4.61 & 23.61\\
8 & 9.22 & 28.22\\
\bottomrule
\relax
\end{tabular}
\par\medskip
\begin{tabular}{lrrrr}
\toprule
\multicolumn{5}{l}{\textbf{b, Held-out prediction}}\\
Layers & \shortstack{Token cross-\\entropy} & \shortstack{Composition\\match (\%)} & \shortstack{$a$ MAE\\(\AA)} & \shortstack{Wyckoff\\parameter MAE}\\\midrule
2 & $0.679\pm0.005$ & $97.36\pm0.08$ & $0.159\pm0.004$ & $0.0960\pm0.0002$\\
4 & $0.637\pm0.003$ & $97.62\pm0.12$ & $0.143\pm0.007$ & $0.0853\pm0.0023$\\
8 & $0.605\pm0.003$ & $97.93\pm0.06$ & $0.137\pm0.008$ & $0.0781\pm0.0006$\\
\bottomrule
\relax
\end{tabular}
\end{table}

Each checkpoint requests 16,384 draws per generation task with one common generation seed. Invalid draws remain in the denominator, and formula deduplication is within each run. The fixed four-layer reference uses the first 16,384 requests of the corresponding archived run without selection by outcome. All latent models independently encode the same 144, 512 and 196 bank structures for the Nd, Fe--Co and Mn tasks (Table~\ref{SI-tabS:aligned-campaigns}). Standardized diagonal GMMs use $K=64$, fitted component weights and $\alpha=1.0$; decoding uses $\tau=1.2$ and top-$k=16$. Geometry, chemistry, frozen ALIGNN screening and the reference of 804,461 formulas match Note~\ref{SI-sec:supp-aligned-controls}. Reported uncertainty is the sample standard deviation across training instances, not repeated generation seeds.

To distinguish reference novelty from within-run repetition, let $H_n$ count screening passes before novelty filtering. The mean reference-unseen fraction $A_n/H_n$, calculated within each run before averaging, decreases from four to eight layers from 10.58\% to 7.92\% for Nd, 2.09\% to 1.70\% for Fe--Co and 7.09\% to 5.99\% for Mn. The distinct fraction $D_n/A_n$ within qualifying reference-unseen draws instead increases in all three tasks. Thus, the lower eight-layer discovery count is associated with fewer qualifying reference-unseen occurrences, rather than increased repetition within that subset.
\FloatBarrier

\subsection{Decoder conditioning pathways}\label{SI-sec:supp-decoder-paths}

The four-layer controls in Table~\ref{SI-tabS:decoder-paths} separate the latent-to-decoder route in Eqs.~\eqref{SI-eq:supp-latent-memory}--\eqref{SI-eq:supp-cross-residual} from direct magnetic-condition modulation in Eq.~\eqref{SI-eq:supp-film}. Full is the same four-layer Signed-CE\&Aux reference used above. Cross-attention only omits the FiLM transformation, so $H^{(l)}=\widetilde H^{(l)}$, while retaining the magnetic condition in the posterior and prior. FiLM only removes the encoder, posterior, prior, latent-to-memory projection and cross-attention; thus $C^{(l)}=S^{(l)}$ and there is no latent sampling or KL term. Unconditioned additionally removes FiLM and receives no external target condition. All four models retain causal self-attention, signed magnetic-moment tokens in decoder history and outputs, and the same token, sign, signed-value and cell-summary losses. The KL term is retained only when a latent distribution exists.

The new variants are trained from random initialization at the same update budget; common surviving modules share their same-seed initial values. Generation follows the protocol above. Cross-attention only uses its own GMM fitted to the common bank members. FiLM only receives magnetic-condition bin 9 for both high-moment tasks and bin 0 for the Mn low-net-moment task, without a chemistry-specific input; Unconditioned receives neither a bank nor a magnetic condition. Neither uses element forcing. These comparisons therefore assess complete conditioning routes, rather than an isolated cross-attention operator with all other modules held fixed.

\begin{table}[!htbp]
\centering\small
\caption{Decoder conditioning pathways and generation under matched budgets. a, Retained input routes and total trainable parameters (millions). b, Screening pass rate before novelty filtering (\%) and distinct qualifying reference-unseen formulas $D_n$ from 16,384 requests. Values are mean $\pm$ sample standard deviation across three training instances with one common generation seed. Full reuses the four-layer baseline in main-text Fig.~\ref{fig:decoder-depth}.}
\label{SI-tabS:decoder-paths}
\setlength{\tabcolsep}{4pt}
\begin{tabular}{lccr}
\toprule
\multicolumn{4}{l}{\textbf{a, Conditioning routes}}\\
Model & Latent route & Direct FiLM & Total parameters\\\midrule
Full & Yes & Yes & 23.61\\
Cross-attention only & Yes & No & 23.21\\
FiLM only & No & Yes & 15.80\\
Unconditioned & No & No & 15.40\\
\bottomrule
\relax
\end{tabular}
\par\medskip
\setlength{\tabcolsep}{3pt}
\resizebox{\textwidth}{!}{%
\begin{tabular}{lrrrrrr}
\toprule
\multicolumn{7}{l}{\textbf{b, Generation results}}\\
& \multicolumn{2}{c}{Nd target} & \multicolumn{2}{c}{Fe--Co target} & \multicolumn{2}{c}{Mn target}\\
\cmidrule(lr){2-3}\cmidrule(lr){4-5}\cmidrule(lr){6-7}
Model & Pass (\%) & $D_n$ & Pass (\%) & $D_n$ & Pass (\%) & $D_n$\\\midrule
Full & $16.39\pm1.72$ & $89.7\pm13.3$ & $18.56\pm1.41$ & $43.7\pm4.0$ & $32.79\pm2.39$ & $361.7\pm39.5$\\
Cross-attention only & $17.41\pm2.31$ & $82.3\pm18.8$ & $19.44\pm0.46$ & $44.7\pm9.8$ & $31.48\pm1.76$ & $297.0\pm53.6$\\
FiLM only & $0.20\pm0.06$ & $5.0\pm2.6$ & $0.23\pm0.08$ & $4.3\pm3.5$ & $1.66\pm0.43$ & $11.7\pm2.1$\\
Unconditioned & $0.006\pm0.000$ & $0.0\pm0.0$ & $0.014\pm0.009$ & $0.7\pm0.6$ & $2.28\pm0.31$ & $13.3\pm0.6$\\
\bottomrule
\relax
\end{tabular}}
\end{table}
\relax
\relax
\section{Retargeting with selected posterior latent vectors}\label{SI-sec:supp-retargeting}

\subsection{Bank selection}

Table~\ref{SI-tabS:posterior-support} distinguishes the banks used for production, strict retargeting and full-training controls. Membership uses stored DFT-derived or surrogate-predicted labels; generated structures are screened by the frozen ALIGNN predictor. Each selected training-structure bank supplies posterior-mean latent vectors; it does not guarantee control over chemistry absent from the bank.

\begin{table}[!htbp]
\centering\small
\caption{Selected training-structure banks and task-specific magnetic criteria. Moments are in $\mu_B$/atom. All banks exclude actinides. Nd requires Nd and excludes Mn; Fe--Co requires Fe and Co and excludes rare earths; Mn requires Mn. The production bank uses conditions 8 or 9, whereas production screening requires $m_{\rm net}\geq1.8$. Strict retargeting additionally constrains aggregates of local magnetic moments and their signs. Full-training controls use only the stored global magnetic-moment label and include nonmagnetic structures in the Mn task. Equal bank sizes across protocols do not establish identical membership.}
\label{SI-tabS:posterior-support}\label{SI-tabS:aligned-campaigns}
\setlength{\tabcolsep}{3pt}
\begin{tabular}{p{0.20\textwidth}p{0.13\textwidth}p{0.49\textwidth}r}
\toprule
Protocol & Target & Magnetic selection & Vectors\\\midrule
Production & Nd & Conditions 8 or 9; no moment-ratio requirement & 394\\
Strict retargeting & Nd & $m_{\rm net}\geq1.8$, $r_{\rm align}\geq0.9$ & 144\\
& Fe--Co & $m_{\rm net}\geq1.8$, $r_{\rm align}\geq0.9$ & 512\\
& Mn & $m_{\rm abs}\geq1.5$, $m_{\rm net}\leq0.1$; both signs beyond $\pm0.1~\mu_B$ & 196\\
Full-training controls & Nd & Global label $\geq1.8$ & 144\\
& Fe--Co & Global label $\geq1.8$ & 512\\
& Mn & Global label $\leq0.1$ & 196\\\bottomrule
\end{tabular}
\end{table}

The production bank samples at most 50,000 training records per original condition before chemistry filtering. Strict banks use reproducibly sampled subsets capped at 512 structures, with equal sizes for chemistry-only, magnetic-only and joint rules within each generation task. Original encoder conditions are retained. Diagonal GMMs use variance regularization $10^{-4}$, three expectation--maximization initializations and at most 500 iterations. Production and response-grid mixtures are fitted separately even when they share a bank. A production-bank scan favours $K=16$ by the Bayesian information criterion \cite{schwarz1978bic}; this density-fit diagnostic does not evaluate generation.

\subsection{Generation and counting}

Retargeting samplers share the frozen generator, geometry test, target and scorer. Five runs per sampler request 2,048 draws at $\tau=0.8$, top-$k=16$ and fitted-bank scale $\alpha=1.0$. Decoder conditions are 9 for high-moment targets and 0 for compensation. Assembly of a periodic crystal from the generated descriptors and the 0.5~\AA{} geometry-tolerance check precede chemistry and magnetic scoring. Direct posterior controls uniformly select a joint-bank member, using $z=\mu_q(x_i,c_i)$ or $z=\mu_q(x_i,c_i)+\exp[\ell_q(x_i,c_i)/2]\odot\epsilon$, where $\ell_q$ is posterior log-variance and $\epsilon\sim\mathcal N(0,I)$; the original encoder condition is retained. Member selection, latent noise and decoding use separate random-number streams, so runs with the same seed index are not paired attempt by attempt.

Most historical reconstruction updates used posterior means because quick validation did not restore training mode; the KL term remained active. Aggregate-posterior resampling is therefore a diagnostic for this checkpoint. The full-training controls restore training mode after validation.

Posterior-mean resampling adds no latent noise but retains stochastic token decoding. For the Nd, Fe--Co and Mn targets, respectively, its screening pass rates are $(18.91\pm0.59)$\%, $(42.84\pm1.03)$\% and $(21.19\pm0.45)$\%, and its distinct qualifying reference-unseen formula counts $D$ are $15.2\pm1.5$, $1.8\pm0.8$ and $1.2\pm0.8$. Values are means $\pm$ sample standard deviations across the five generation seeds. This control has the highest screening pass rates among the tested samplers; its distinct-formula yield matches Joint GMM for Nd but is lower for Fe--Co and Mn (main-text Table~\ref{tab:retargeting}).

Qualification indicator $H_j$ records passing decoding and geometry checks and meeting the chemical and magnetic criteria specified for the task and experimental protocol, without a novelty requirement. For reference-novelty indicator $U_j=\mathbf1\{F_j\notin\mathcal R\}$, set $I_j=H_jU_j$; undecodable attempts have $H_j=U_j=0$. The reference contains 804,461 reduced formulas. With $H_n=\sum_{j=1}^{n}H_j$, the screening pass rate is $H_n/n$. The joint probability $Q=\Pr(I_j=1)=\Pr(H_j=1,U_j=1)$ includes both qualification and reference novelty. The count $A_n$ of qualifying reference-unseen draws and their distinct-formula count $D_n$ follow main-text Eqs.~\MainEq{eq:yield-counts}--\MainEq{eq:distinct-count}. The distinct qualifying-formula count without a novelty requirement is $J_n=|\{F_j:H_j=1\}|$, and the repeated-formula fraction among qualifying reference-unseen draws is $1-D_n/A_n$, undefined at $A_n=0$. A formula enters a set at its first qualifying occurrence, even if an earlier draw failed screening. Invalid requests remain in denominators; cumulative checkpoints are dependent, and deduplication is within a run.

The Source Data retain all eight samplers, individual runs and threshold sensitivity. Four geometry-passing requests contain unsupported superheavy species and have unresolved magnetic scores, but independently fail chemistry; their joint indicators are zero. Affected magnetic-only pass rates remain undefined without changing joint or novelty counts.

\subsection{Latent-space display}

For main-text Fig.~\ref{fig:samplers}, pooled principal components and plotting limits are fixed across panels. Projected histogram $H^{(s)}$ is Gaussian-smoothed to $\widetilde H^{(s)}$ and displayed as
\begin{equation}
R^{(s)}_{pq}=\frac{\widetilde H^{(s)}_{pq}}{\max_{p',q'}\widetilde H^{(s)}_{p'q'}}.
\label{SI-eq:projected-density}
\end{equation}
This normalization compares spatial distributions, not absolute density magnitudes.
\FloatBarrier
\relax
\section{Finite-budget yield}\label{SI-sec:supp-note-3}

\subsection{Distinct-formula accumulation and response validation}

The repeat-inclusive count $A_n$ of qualifying reference-unseen draws and their distinct-formula count $D_n$ use main-text Eqs.~\MainEq{eq:yield-counts}--\MainEq{eq:distinct-count}. For a fixed sampling setting with stationary formula probabilities, let $p_f=\Pr(F=f\mid I=1)$. Independent draws from this qualifying reference-unseen subset then give
\begin{equation}
\mathbb E[D(a)]=\sum_f\left[1-(1-p_f)^a\right].
\label{SI-eq:exact-occupancy}
\end{equation}
This statistical occupancy model describes how qualifying reference-unseen draws populate formula categories, distinct from the atomic site occupancy considered in Note~\ref{SI-sec:supp-note-4}. Because short generation runs cannot reliably estimate all $p_f$, main-text Eq.~\MainEq{eq:occupancy-law} approximates finite-budget distinct-formula accumulation within the tested workflow. Its late-accumulation exponent is not an infinite-space law: for a finite admissible formula set, the exact logarithmic slope ultimately tends to zero.

The grid combines $K\in\{32,64,128\}$, $\alpha\in\{0.5,0.75,1.0,1.25,1.5\}$ and $\tau\in\{0.8,1.0,1.2\}$: 45 sampling settings with 50,000 draws each and 56 cumulative checkpoints from 100 to 50,000 draws. Latent random-number streams are shared across $\tau$ at fixed $(K,\alpha)$; checkpoints are repeated observations within generation runs rather than independent replicates. Records retain sampling parameters, component, latent vector, decoding and screen outcomes, reduced formula and novelty status. Through-origin fits to $\mathbb E[A_n]=nQ$ have median uncentred $R_0^2=0.999$ and minimum 0.994, while the median qualifying reference-unseen draw rate falls 1.6\% between the two 25,000-draw halves.

The fit to the joint probability $Q$ uses a product-binomial composite negative log-likelihood, an $L_2$ penalty of $10^{-3}$ on non-intercept coefficients, analytic gradients and the limited-memory Broyden--Fletcher--Goldfarb--Shanno algorithm with bounds (L-BFGS-B)~\cite{byrd1995lbfgsb}. The accumulation model uses trust-region-reflective least squares with soft-$L_1$ scale 0.12. Global starts span $\beta_{\rm late}\in\{0.55,0.65,0.75\}$, $a_0\in\{3,8,20\}$ and $\gamma\in\{1,2,4\}$; the converged solution minimizing mean squared log residual is retained. Comparator fits for held-out $(\alpha,\tau)$ pairs fix training-fold $\gamma$ and use nine starts from $\beta_{\rm late}\in\{0.50,0.65,0.80\}$ and $a_0\in\{2,6,20\}$.

The GMM weight-and-covariance descriptor in main-text Eq.~\MainEq{eq:gmm-descriptor} combines weights and covariance volumes before scaling the component spread. Its three values correspond one-to-one to $K$, so their associations are not separable. Five-fold sampling-setting holdouts for Eq.~\MainEq{eq:qualifying-response} give MAE $8.3\times10^{-4}$, MAPE 8.73\%, Pearson $r=0.948$ and predictive $R^2=0.896$. With the observed qualifying reference-unseen draw count supplied, accumulation MAPE is 11.2\%; combining the cross-validated probability and accumulation models gives 12.6\% MAPE and endpoint correlation 0.600. These tests assess recorded-grid interpolation.

Table~\ref{SI-tabS:occupancy-selection} uses repeated short-grid folds to compare response specifications and characterize interpolation; separate long generation runs provide post-fit checks. Accumulation RMSE and $K$-bias condition on the observed qualifying reference-unseen draw count; endpoint errors use the combined rate--accumulation prediction. Each held-out $(\alpha,\tau)$ pair supplies one fitted $(\beta_{\rm late},a_0)$ pair across its three runs with different $K$ at training-only $\gamma$. Response predictions average $\beta_{\rm late}$ arithmetically and $a_0$ geometrically across $K$; plotted $a_0$ values are parameter-pair summaries.

\begin{table}[t]
\centering
\caption{Response specifications for distinct-formula accumulation. Each specification fits one shared $\ln\gamma$; covariates are defined below. Values are medians over five deterministic repetitions of folds withholding entire sampling settings or $(\alpha,\tau)$ pairs. Accumulation RMSE is computed on $\ln(1+D)$; $K$-bias is the root mean square across $K$ of the $K$-specific mean $\ln(1+D)$ residual. The final column reports two ten-million-draw post-fit checks that were not used to fit the coefficients. Lower is better.}
\label{SI-tabS:occupancy-selection}
\small
\setlength{\tabcolsep}{2.5pt}
\begin{tabular}{@{}lrrrrr@{}}
\toprule
Response & \shortstack{Coeff.\\count} & \shortstack{Sampling-setting\\RMSE} & \shortstack{50k MAPE\\(\%)} & \shortstack{$K$-bias\\RMS} & \shortstack{10M error\\(\%)}\\
\midrule
Symmetric & 7 & 0.184 & 19.88 & 0.208 & 17.65\\
Reduced & 6 & 0.182 & 18.68 & 0.202 & 18.92\\
Additive descriptor & 8 & 0.128 & 11.81 & 0.046 & 6.28\\
Selected interaction & 8 & 0.123 & 11.94 & 0.040 & 7.08\\
\bottomrule
\end{tabular}
\end{table}

With $C=\mathcal C(\widehat\Theta_K)$, the covariate sets for $(\operatorname{logit}\beta_{\rm late};\ln a_0)$ are $(\{\alpha,\tau\};\{\alpha,\tau\})$ for the symmetric response, $(\{\tau\};\{\alpha,\tau\})$ for the reduced response, $(\{\alpha,\tau\};\{\alpha,\tau,C\})$ for the additive descriptor response and $(\{\tau\};\{\alpha,\tau,C,\alpha C\})$ for the selected interaction. Each covariate has its own coefficient, and each linear predictor includes an intercept.

For the selected response, $\partial\ln a_0/\partial\alpha=d_\alpha+d_{\alpha C}\mathcal C(\widehat\Theta_K)$. The interaction is negative in all 25 fits with held-out parameter pairs, but its interpretation remains conditional on the three mixtures. Across five repeated parameter-pair holdouts, median MAE/Pearson $r$ are 0.056/0.435 for $\beta_{\rm late}$ and 4.81/0.416 for $a_0$; fitted shared sharpness is $\gamma=1.487$. The 1,000 parameter-pair resampling refits quantify fitted-grid sensitivity, not confidence intervals or independent generator variation. No null-hypothesis tests are used.

\subsection{Scale interpretation and separate long generation runs}

For standardized draw $u_j\mid k_j\sim\mathcal N(\widehat\mu_{k_j},\alpha^2\widehat\Sigma_{k_j})$ from component $k_j$, the normalized squared Mahalanobis radius is
\begin{equation}
\rho_j^2=\frac{(u_j-\widehat\mu_{k_j})^{\mathsf T}\widehat\Sigma_{k_j}^{-1}(u_j-\widehat\mu_{k_j})}{d}.
\end{equation}
Writing $u_j=\widehat\mu_{k_j}+\alpha L_{k_j}\epsilon_j$, $L_{k_j}L_{k_j}^{\mathsf T}=\widehat\Sigma_{k_j}$ and $\epsilon_j\sim\mathcal N(0,I_d)$ yields
\begin{align}
\rho_j^2&\sim\frac{\alpha^2}{d}\chi_d^2,\\
\rho_{\rm rms}&=\sqrt{\mathbb E[\rho_j^2]}=\alpha.
\label{SI-eq:supp-alpha-radius}
\end{align}
Thus $\alpha$ and $\rho_{\rm rms}$ are the same sampler covariate, not independent predictors.

Coefficients are fitted only to the 45 short generation runs. The separate $(K,\alpha,\tau)=(64,1.0,1.2)$ ten-million-draw run yields 3,334 formulas against 3,576 predicted (7.3\% error), with observed/predicted qualifying reference-unseen draw rates 0.85\%/0.82\%. The out-of-range $K=16$ run at $(16,0.5,0.8)$ yields 1,689 against 1,572 ($-6.9\%$), despite overprediction of $Q$ (1.00\%/1.72\%). 

Each long test comprises forty 250,000-draw blocks, recording cumulative counts every 50 draws and each qualifying reference-unseen formula, component and latent vector. Block offsets and within-block indices restore global order. Intermediate observed bands permute block order without changing endpoints. Model-sensitivity ranges use 1,000 resamples of the 15 $(\alpha,\tau)$ pairs, retaining all $K$ values for each pair.

\subsection{Finite-budget numerical evaluation}

Ordinary composed predictions use a stable binomial sum with omitted tail probability below $10^{-12}$. For arrays containing more than 10,000 predictions, entries with $\bar a=n\widehat Q\geq64$ use the fourth-order central-moment expansion; entries below this mean retain the binomial sum. Writing $f(a)=\mathcal D(a)$ and $v=n\widehat Q(1-\widehat Q)$, the approximation is
\begin{equation}
\begin{aligned}
\mathbb E[f(A_n)]\simeq{}&f(\bar a)+\frac v2f''(\bar a)
+\frac{v(1-2\widehat Q)}6f'''(\bar a)\\
&+\frac{3v^2+v[1-6\widehat Q(1-\widehat Q)]}{24}f^{(4)}(\bar a).
\end{aligned}
\end{equation}
The implementation clips the result to $[0,n\widehat Q]$ and returns zero at zero mean. Comparison with direct sums at 2,000 archived long-horizon prediction points gave a maximum relative difference below $6.2\times10^{-8}$; a fixed 576-case check spanning the crossover and the threshold in $n\widehat Q$ gave less than $9.7\times10^{-7}$. Inputs, exact and approximate predictions, and errors are supplied in the Source Data; these checks do not refit the response.

\FloatBarrier
\relax
\section{Magnetic ablation experiments}\label{SI-sec:supp-aligned-controls}

\subsection{Model interventions and training}

The five models with site-resolved magnetic-moment outputs isolate auxiliary supervision or signed input paths (Table~\ref{SI-tabS:aligned-design}); Global-GReg provides a global-only system benchmark. Signed-CE\&Aux retrains the primary model configuration from random initialization and is distinct from the production checkpoint. Absolute-input models retain signed token targets and output distributions, so they test input-path access to signs rather than removing sign information from training.

\begin{table}[!htbp]
\centering
\caption{Information paths and supervision in controls evaluated under matched conditions. Every model receives the same ten-bin global magnetic condition. ``Signed'' and ``absolute'' describe local magnetic-moment inputs; a slash denotes an absent local magnetic-moment channel. Auxiliary losses are the additional sign, signed-value and cell-summary terms. A direct global output uses a terminal continuous scalar head and its SmoothL1 loss. Initialization differences are described below.}
\label{SI-tabS:aligned-design}
\resizebox{\textwidth}{!}{%
\begin{tabular}{llllll}
\toprule
Model & Encoder moments & History moments & Moment output & Aux. losses & Global head/loss\\
\midrule
Signed-CE\&Aux & Signed & Signed & Signed & On & No\\
Signed-CE & Signed & Signed & Signed & Off & No\\
EncAbs-CE & Absolute & Signed & Signed & Off & No\\
HistAbs-CE & Signed & Absolute & Signed & Off & No\\
DualAbs-CE & Absolute & Absolute & Signed & Off & No\\
Global-GReg & / & / & / & Off & Yes\\
\bottomrule
\end{tabular}}
\end{table}

EncAbs-CE takes absolute encoder moments, HistAbs-CE maps previous moments to absolute-value tokens during teacher forcing and autoregression, and DualAbs-CE applies both. All five site-output models retain signed targets and outputs, reliability weighting, masks, 202 positions and 23,613,098 parameters. Signed-CE removes the three magnetic auxiliary losses from Signed-CE\&Aux. The three absolute-input models additionally change the designated inputs relative to Signed-CE; token-loss normalization is unchanged.

Global-GReg receives no individual site magnetic moments and has no site-moment tokens, outputs, targets or locally derived summary losses. Its terminal $256\to128\to1$ multilayer perceptron uses GELU and a nonnegative Softplus output for the stored scalar total-moment label divided by the cell atom count; only its ten-bin condition, not the continuous target value, is an input. With mean active structural-token cross-entropy $\mathcal L_{\rm struct}$, its objective is
\begin{equation}
\mathcal L_{\text{Global-GReg}}=\mathcal L_{\rm struct}+\mathcal L_{\rm global}+10^{-3}\mathcal L_{\rm KL},
\end{equation}
where $\mathcal L_{\rm global}$ is SmoothL1 with transition $1~\mu_B$/atom. Its 170 positions, 22,196,473 parameters and named-module initialization differ from the site-output family. The scalar labels are retained from the existing dataset, including previously predicted values; they are not recomputed from site moments during this control.

All configurations use three seeds and final checkpoints at 560,000 updates. Within the five-model family, same-seed initial weights and data-sampling streams are identical, with validation randomness isolated uniformly. Table~\ref{SI-tabS:model-settings} gives the shared optimization settings. Global-GReg is a system-level benchmark because its architecture, objective and initialization differ.

\subsection{Common generation tasks and distinct-formula discovery}

The 18 checkpoints, three generation tasks and three generation seeds give 162 runs of 100,000 requests (16.2 million draws). Checkpoints encode identical selected training structures under their original conditions, then independently fit standardized GMMs. Bank selection uses chemistry and the stored global magnetic-moment label alone (Table~\ref{SI-tabS:aligned-campaigns}). Oversubscribed banks retain the smallest SHA-256 hashes of a fixed seed, generation task and record identifier, ordered by dataset index.

The $K=64$ GMMs use covariance regularization $10^{-4}$, three initializations, at most 500 iterations and a common fitting seed. Methods specify the common decoding settings. Generation uses 256-request chunks and scorer batches of 32. The frozen ALIGNN receives only the generated crystal. A generation attempt passes screening if it satisfies assembly, geometry, task-specific chemistry and the frozen ALIGNN global-moment criterion. Invalid requests, including zero-volume cells, remain in the denominator without replacement. Interrupted decoding is resumed from the original sampled tokens; zero-volume cells are marked invalid rather than repaired or resampled.

\begin{figure}[!htbp]
\centering
\includegraphics[width=\textwidth]{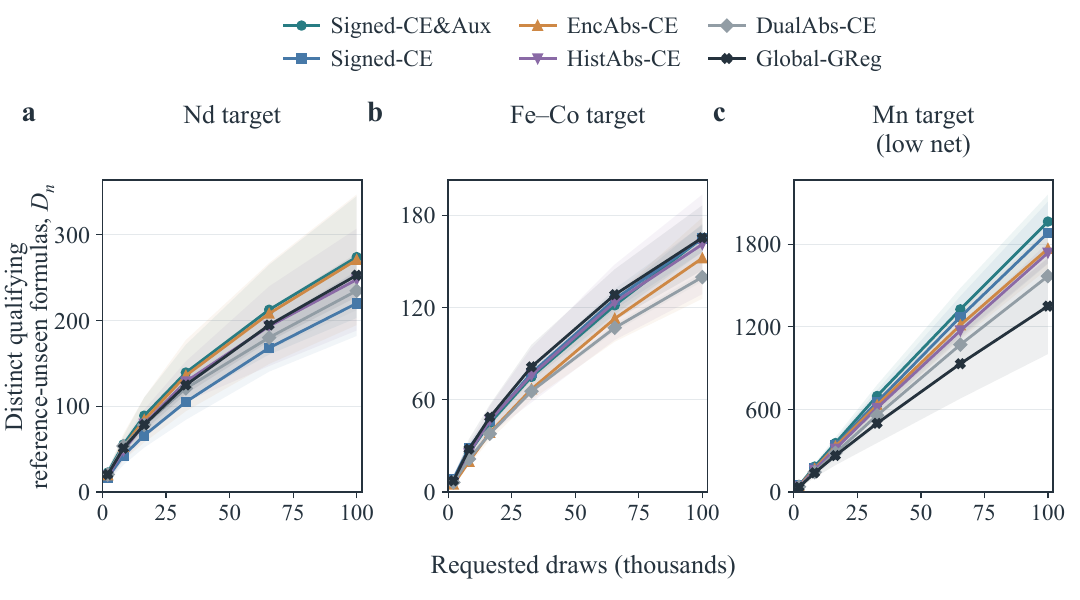}
\caption{Accumulation of different qualifying reference-unseen formulas in the generation experiment under matched conditions. Panels show the Nd-containing high-moment, rare-earth-free Fe--Co high-moment and Mn-containing low-net-moment generation tasks, respectively. At each requested-draw budget, the three generation-seed curves are averaged within each training seed; lines and shaded bands show the mean and sample standard deviation across the three training seeds. Each curve deduplicates qualifying reference-unseen formulas within its own run.}
\label{SI-figS:aligned-accumulation}
\end{figure}

Generation repeats are averaged within training seed/task before reporting training-seed mean and sample standard deviation. Table~\ref{SI-tabS:aligned-quality} separates reference novelty from within-run repetition; Fig.~\ref{SI-figS:aligned-accumulation} shows accumulation. Individual runs, geometry rates and 300,000-request formula unions across each checkpoint's three generation seeds accompany the Source Data.

\begin{table}[!htbp]
\centering\small
\caption{Formula novelty and repetition under matched generation conditions. The reference-unseen fraction among screening passes is $A_n/H_n$, where $H_n$ counts passes before novelty filtering; the distinct-formula fraction among qualifying reference-unseen draws is $D_n/A_n$. Ratios pool counts within each model/task across runs, with $D_n$ deduplicated within each run. Both are percentages and have different denominators from the per-request discovery count.}
\label{SI-tabS:aligned-quality}
\setlength{\tabcolsep}{3pt}
\begin{tabular}{lrrrrrr}
\toprule
& \multicolumn{2}{c}{Nd task} & \multicolumn{2}{c}{Fe--Co task} & \multicolumn{2}{c}{Mn task}\\
\cmidrule(lr){2-3}\cmidrule(lr){4-5}\cmidrule(lr){6-7}
Model & $A_n/H_n$ & $D_n/A_n$ & $A_n/H_n$ & $D_n/A_n$ & $A_n/H_n$ & $D_n/A_n$\\\midrule
Signed-CE\&Aux & 10.14 & 16.49 & 2.04 & 43.63 & 7.08 & 84.64 \\
Signed-CE & 10.16 & 14.22 & 2.12 & 41.76 & 6.67 & 87.74 \\
EncAbs-CE & 10.42 & 15.32 & 1.58 & 47.17 & 6.30 & 83.44 \\
HistAbs-CE & 9.20 & 16.69 & 1.89 & 44.95 & 6.29 & 86.72 \\
DualAbs-CE & 9.71 & 14.54 & 1.48 & 44.06 & 5.91 & 83.28 \\
Global-GReg & 10.34 & 16.53 & 1.81 & 41.03 & 5.58 & 80.88 \\
\bottomrule
\relax
\end{tabular}
\end{table}

\subsection{Magnetic-moment sign agreement and sensitivity}

Exact magnetic-moment sign agreement compares each site's generated moment sign with the frozen ALIGNN prediction on the same generated structure and requires a match at every site with a nonzero ALIGNN-predicted moment. Each model is evaluated on its own generated structures. Global-GReg has no corresponding metric. The audit samples the same 16 complete 256-request chunks without replacement from indices 0--389 in all 135 site-output runs, divided into four disjoint 1,024-request batches. This gives 4,096 requests per run and 552,960 overall; the final 160 requests are uniformly outside the frame. Selection does not depend on hits, novelty or scores. Exact indices are supplied in the Source Data.

Original signed tokens are expanded over crystallographic orbits, including for HistAbs-CE and DualAbs-CE. A unique element-preserving bijection matches archived scorer atoms within 0.001~\AA{} periodic discrepancy and requires matching lattices. Eligibility requires decoding, geometry and task-specific chemistry, without a magnetic-hit or novelty condition. For generated moments $g_i$ and frozen ALIGNN reference moments $r_i$, one global sign orientation minimizes the mean absolute moment error over all atoms
\begin{equation}
\eta=\operatorname*{arg\,min}_{h\in\{-1,+1\}}\frac{1}{N}\sum_{i=1}^{N}|h g_i-r_i|,
\end{equation}
with $+1$ retained at ties. For $s(x)=+1,0,-1$ above $0.1~\mu_B$, within $[-0.1,0.1]~\mu_B$, and below $-0.1~\mu_B$, respectively, a mixed-sign reference means that the ALIGNN predictions contain both positive and negative nonzero moments. Exact magnetic-moment sign agreement on this subset is
\begin{equation}
E=\prod_{i:\,s(r_i)\ne0}\mathbf{1}\!\left[s(\eta g_i)=s(r_i)\right].
\end{equation}
No site-wise flips are permitted. Reference zeros are excluded from the sign test but remain in the orientation objective; any wrong or generated-zero sign at a reference-nonzero site gives $E=0$. The sign test does not measure moment magnitudes.

Each run pools the four batches' numerators and denominators before averaging generation seeds within training seed/task, then generation tasks equally, then the three training seeds. Each generated occurrence in the mixed-sign reference subset receives equal weight within its run; repeated formulas retain their generated frequencies without altering task weights. Empty subsets are undefined, and aggregates requiring missing constituents are not silently averaged over the remainder. The number of eligible occurrences with mixed reference signs is the denominator (Table~\ref{SI-tabS:aligned-sign-campaigns}).

Weighting formulas equally first averages $E$ within each formula/run, then weights formulas equally under the same outer hierarchy. Site-count strata are 2, 3--4, 5--8, 9--16 and at least 17 nonzero reference sites. The four means calculated separately for the disjoint subsamples assess sampling sensitivity, not independent training replication or confidence intervals (Table~\ref{SI-tabS:aligned-sign-sensitivity}).

Signed-moment MAE first averages $|\eta g_i-r_i|$ over atoms, then eligible structures within a run. Mixed-sign reference fraction is the fraction of geometry- and chemistry-eligible structures with both reference signs. Neither diagnostic adds a filter to the exact magnetic-moment sign-agreement subset.

\begingroup
\small
\setlength{\tabcolsep}{3pt}
\begin{table}[!htbp]
\centering
\caption{Magnetic-moment agreement by generation task. Exact magnetic-moment sign agreement requires an eligible structure with both nonzero ALIGNN-predicted moment signs; the mixed-sign fraction and MAE use all geometry- and chemistry-eligible structures and all constituent elements. Values are training-seed mean $\pm$ sample standard deviation after averaging generation repeats. $N_{\rm mixed}$ counts eligible occurrences with mixed reference signs across the nine runs.}\label{SI-tabS:aligned-sign-campaigns}
\begin{tabular}{llrrrr}
\toprule
Model & Task & \shortstack{Exact sign\\agreement (\%)} & \shortstack{Mixed-sign structures\\(\%)} & \shortstack{MAE\\($\mu_B$)} & $N_{\rm mixed}$\\
\midrule
Signed-CE\&Aux & Nd & $87.59 \pm 2.89$ & $51.55 \pm 0.51$ & $0.65 \pm 0.02$ & 6895 \\
Signed-CE\&Aux & Fe--Co & $0.68 \pm 0.42$ & $10.64 \pm 0.49$ & $1.08 \pm 0.03$ & 2196 \\
Signed-CE\&Aux & Mn (low net) & $23.55 \pm 3.69$ & $20.50 \pm 1.29$ & $0.37 \pm 0.02$ & 3829 \\
Signed-CE & Nd & $87.84 \pm 0.76$ & $50.58 \pm 2.20$ & $0.73 \pm 0.03$ & 6824 \\
Signed-CE & Fe--Co & $0.71 \pm 0.45$ & $11.64 \pm 0.40$ & $1.08 \pm 0.08$ & 2382 \\
Signed-CE & Mn (low net) & $15.48 \pm 5.33$ & $20.40 \pm 3.16$ & $0.37 \pm 0.03$ & 3697 \\
EncAbs-CE & Nd & $85.14 \pm 0.45$ & $51.84 \pm 3.95$ & $0.72 \pm 0.06$ & 7573 \\
EncAbs-CE & Fe--Co & $3.07 \pm 1.08$ & $11.09 \pm 0.40$ & $1.04 \pm 0.02$ & 2296 \\
EncAbs-CE & Mn (low net) & $11.85 \pm 2.90$ & $19.93 \pm 0.22$ & $0.37 \pm 0.02$ & 3829 \\
HistAbs-CE & Nd & $86.21 \pm 1.36$ & $48.57 \pm 3.32$ & $0.69 \pm 0.06$ & 7007 \\
HistAbs-CE & Fe--Co & $0.28 \pm 0.18$ & $11.06 \pm 0.63$ & $1.06 \pm 0.02$ & 2232 \\
HistAbs-CE & Mn (low net) & $17.63 \pm 3.95$ & $21.32 \pm 0.49$ & $0.39 \pm 0.02$ & 3996 \\
DualAbs-CE & Nd & $85.99 \pm 0.68$ & $49.59 \pm 3.99$ & $0.71 \pm 0.04$ & 6712 \\
DualAbs-CE & Fe--Co & $2.02 \pm 0.55$ & $10.71 \pm 0.79$ & $1.03 \pm 0.06$ & 2226 \\
DualAbs-CE & Mn (low net) & $5.39 \pm 2.20$ & $19.80 \pm 2.26$ & $0.38 \pm 0.03$ & 3722 \\
\bottomrule
\relax
\end{tabular}
\end{table}
\endgroup

\begin{table}[!htbp]
\centering\small
\caption{Sensitivity of exact magnetic-moment sign agreement. a, Equal weighting per generated occurrence or per formula and the range of four disjoint batch means. b, Groups defined by the number of sites with nonzero ALIGNN-predicted moments: agreement (\%) and pooled eligible-sample count $n$. Values use training-seed mean $\pm$ sample standard deviation and equal generation-task weights. NA marks an undefined aggregate when a required averaging group has no eligible samples.}
\label{SI-tabS:aligned-sign-sensitivity}\label{SI-tabS:aligned-sign-size}
\setlength{\tabcolsep}{3pt}
\begin{tabular}{lrrr}
\toprule
\multicolumn{4}{l}{\textbf{a, Weighting and subsampling}}\\
Model & \shortstack{Equal per occurrence\\agreement (\%)} & \shortstack{Equal per formula\\agreement (\%)} & \shortstack{Batch range\\(\%)}\\\midrule
Signed-CE\&Aux & $37.27 \pm 0.74$ & $28.37 \pm 1.07$ & 36.59--37.73 \\
Signed-CE & $34.68 \pm 1.52$ & $28.21 \pm 1.07$ & 34.11--35.16 \\
EncAbs-CE & $33.35 \pm 1.31$ & $29.64 \pm 0.93$ & 32.74--34.29 \\
HistAbs-CE & $34.71 \pm 1.21$ & $26.67 \pm 0.95$ & 34.40--35.25 \\
DualAbs-CE & $31.13 \pm 0.49$ & $29.42 \pm 0.19$ & 30.87--31.40 \\
\bottomrule
\relax
\end{tabular}
\medskip
\renewcommand{\arraystretch}{1.18}
\begin{tabular}{lccccc}
\toprule
\multicolumn{6}{l}{\textbf{b, Sites with nonzero ALIGNN-predicted moments}}\\
Model & 2 & 3--4 & 5--8 & 9--16 & $\geq17$\\\midrule
Signed-CE\&Aux & \shortstack{NA\\$n=154$} & \shortstack{$47.25 \pm 7.71$\\$n=1322$} & \shortstack{$39.94 \pm 2.65$\\$n=2658$} & \shortstack{$34.01 \pm 0.61$\\$n=3940$} & \shortstack{$28.50 \pm 0.80$\\$n=4846$} \\
Signed-CE & \shortstack{NA\\$n=220$} & \shortstack{NA\\$n=1238$} & \shortstack{$36.38 \pm 2.79$\\$n=2771$} & \shortstack{$32.00 \pm 1.17$\\$n=3730$} & \shortstack{$30.74 \pm 2.35$\\$n=4944$} \\
EncAbs-CE & \shortstack{NA\\$n=129$} & \shortstack{$36.64 \pm 5.25$\\$n=1329$} & \shortstack{$36.95 \pm 2.45$\\$n=2732$} & \shortstack{$32.39 \pm 1.22$\\$n=3893$} & \shortstack{NA\\$n=5615$} \\
HistAbs-CE & \shortstack{NA\\$n=238$} & \shortstack{$44.72 \pm 4.82$\\$n=1339$} & \shortstack{$35.07 \pm 1.97$\\$n=2680$} & \shortstack{$33.43 \pm 1.66$\\$n=3741$} & \shortstack{$27.82 \pm 0.73$\\$n=5237$} \\
DualAbs-CE & \shortstack{NA\\$n=204$} & \shortstack{NA\\$n=1109$} & \shortstack{$33.78 \pm 0.36$\\$n=2664$} & \shortstack{$31.22 \pm 1.27$\\$n=4068$} & \shortstack{NA\\$n=4615$} \\
\bottomrule
\relax
\end{tabular}
\end{table}

\subsection{Metric selection and interpretation}

Among 265,465 eligible structures, 65,416 have mixed reference signs. Signed-CE\&Aux exceeds Signed-CE in exact magnetic-moment sign agreement with equal weight per generated occurrence in all training seeds. Signed-CE exceeds EncAbs-CE and DualAbs-CE in all seeds but trails HistAbs-CE's mean by 0.03 percentage points. Equal weighting per formula instead places EncAbs-CE and DualAbs-CE above both signed-input models (Table~\ref{SI-tabS:aligned-sign-sensitivity}). The relative ranking therefore depends on the weighting convention.

The diagnostic originated in exploration of 49 related or overlapping magnetic and generation metrics. Four candidate metrics were fixed before a nonoverlapping historical subsample was evaluated; those exploratory values are not substituted into the present results. The current definition and sampling protocol were fixed before auditing this archive of controls evaluated under matched conditions. All five models, three generation tasks, seeds, batches and stated sensitivities are retained regardless of ordering. Because the predictor also supplied surrogate-predicted training labels, agreement measures surrogate consistency, with site count and generated-structure distributions affecting comparisons.
\relax
\FloatBarrier
\begin{samepage}
\section{Physical evaluation}\label{SI-sec:supp-note-4}

\subsection{Nd generation experiment at a fixed sampling configuration}

The illustrative ten-million-draw generation experiment uses the frozen generator and Nd production bank at $(K,\alpha,\tau)=(64,1.0,1.2)$. It is a separate realization from the yield-validation trajectory in Note~\ref{SI-sec:supp-note-3}. Main-text Methods specify DFT selection; Table~\ref{SI-tabS:physical-settings} collects the calculation settings. Non-finite energies, short contacts, large volume excursions, changed species and unusable coordinates are excluded.
\end{samepage}

Novelty classes are mutually exclusive: unseen formula; seen formula/unseen space group; seen formula/seen space group. The generation experiment uses the first class against the complete generator formula reference, fixed before physical evaluation.

Of 36 new DFT calculations, 33 yield usable results and three fail to converge. Historical formula/atom-count matches are checked against eSEN-relaxed candidates by element-preserving structure matching in pymatgen \cite{ong2013pymatgen}, using exact elements, $l_{\rm tol}=s_{\rm tol}=0.1$, $5^\circ$ angular tolerance and cell scaling, without primitive reduction or supercell search. Five of six pass; one with exchanged Nd/Sm occupations is excluded. Table~\ref{SI-tabS:dft-funnel} records these outcomes and overlapping energy/magnetic classifications.

\begin{table}[!htbp]
\centering
\small
\caption{Example ten-million-draw Nd generation experiment. Reference-unseen formula status is fixed against the complete generator reference before eSEN or DFT inspection. The DFT rows distinguish new calculations from historical records confirmed as structure-equivalent; one historical mismatch is excluded. The final three rows are overlapping classifications of the usable structure-level results rather than additional sequential filters.}
\label{SI-tabS:dft-funnel}
\begin{tabular}{p{0.82\textwidth}r}
\toprule
Selection stage or outcome & Count \\
\midrule
Requested latent draws & 10,000,000 \\
Distinct qualifying reference-unseen formulas & 3,320 \\
Archived eSEN convergence within 200 steps & 946 \\
Converged, $E_{\rm form}^{\rm eSEN}<0$ and $M_{\rm pred}/N\geq1.8~\mu_B\,{\rm atom}^{-1}$; selected for DFT & 42 \\
New candidate-specific DFT calculations & 36 \\
Usable results from new calculations & 33 \\
Structure-equivalent historical results & 5 \\
Usable structure-level DFT results & 38 \\
Among the 38: negative DFT formation energy & 34 \\
Among the 38: $\mu_0M_s^{\rm col,est}>1.5$~T from DFT & 26 \\
Among the 38: both DFT criteria & 22 \\
\bottomrule
\end{tabular}
\end{table}

\subsection{Calculation settings}

Candidate and competing-phase relaxations use eSEN-30M-OAM \cite{fu_esen}, pretrained on OMat24 and fine-tuned on MPtrj and sAlexandria data \cite{barroso_omat24,deng2023chgnet,schmidt2024alexandria}. The workflow uses the Atomic Simulation Environment \cite{larsen2017ase}, fast inertial relaxation engine \cite{bitzek2006fire} and Fr\'echet cell filter.

\begin{table}[!htbp]
\centering\small
\caption{Settings for candidate validation, chemical site-occupancy configurations and eSEN relaxation. DFT calculations use spin-polarized PBE; site-occupancy calculations keep the parent cell fixed. The eSEN convex-hull calculation applies the same checkpoint and relaxation protocol to candidates and competing phases.}
\label{SI-tabS:physical-settings}
\renewcommand{\arraystretch}{1.15}
\begin{tabular}{@{}>{\raggedright\arraybackslash}p{0.34\textwidth}>{\raggedright\arraybackslash}p{0.29\textwidth}>{\raggedright\arraybackslash}p{0.29\textwidth}@{}}
\toprule
\multicolumn{3}{l}{\textbf{a, DFT protocols}}\\
Setting & Candidate validation & \shortstack[l]{Site-occupancy\\configurations}\\\midrule
Plane-wave cutoff & 520~eV & 400~eV\\
Reciprocal spacing & $\approx0.20$~\AA$^{-1}$ & 0.30~\AA$^{-1}$\\
Electronic convergence & $10^{-6}$~eV & $10^{-5}$~eV\\
Maximum force & 0.02~eV~\AA$^{-1}$ & 0.04~eV~\AA$^{-1}$\\
Relaxed degrees of freedom & Cell and ions & Ions only\\
Symmetry & Disabled & Disabled\\
\bottomrule
\end{tabular}
\par\medskip
\begin{tabular}{@{}>{\raggedright\arraybackslash}p{0.34\textwidth}>{\raggedright\arraybackslash}p{0.62\textwidth}@{}}
\toprule
\multicolumn{2}{l}{\textbf{b, eSEN relaxation}}\\
Stage & Setting\\\midrule
Initial candidate screening & FIRE/Fr\'echet cell relaxation, at most 200 steps; convergence requires the optimizer flag or final maximum force $\leq0.02$~eV~\AA$^{-1}$\\
Hull pre-relaxation & Fixed-cell FIRE, at most 200 steps, $f_{\max}=0.10$~eV~\AA$^{-1}$\\
Hull full relaxation & FIRE/Fr\'echet, nominal zero pressure, at most 3,000 steps; acceptance requires $f_{\max}\leq0.02$~eV~\AA$^{-1}$ and maximum absolute stress $\leq0.005$~eV~\AA$^{-3}$\\
\bottomrule
\end{tabular}
\end{table}

Candidate validation uses VASP 6.5.0 with PBE64 PAW datasets and first-order Methfessel--Paxton smearing of width 0.10~eV. Cell and ion relaxation allows at most 200 steps. The single collinear initialization assigns 3~$\mu_B$ to rare earths, 5~$\mu_B$ to magnetic $3d$ sites and zero elsewhere, without spin--orbit coupling or competing-state enumeration.

abICS version 2.2.1 varies chemical site occupations on four fixed-composition topologies. Nd$_2$Fe$_{14}$Co$_3$ redistributes 42 Fe and 9 Co over 51 transition-metal sites; NdYFe$_{14}$Co$_3$ and NdSmFe$_{14}$Co$_3$ also redistribute two rare-earth species over six sites; NdFe$_7$Co$_4$ varies Fe/Co occupation. Each system starts with 50 DFT-evaluated configurations and adds 20 in each of three active-learning rounds. The ten linearly spaced temperature replicas span 300--1800~K. Each replica runs 10,000 Monte Carlo steps per round for Nd$_2$Fe$_{14}$Co$_3$ and NdFe$_7$Co$_4$, and 100,000 for the NdY and NdSm systems. Exchange attempts occur every four steps; frames are stored every 16 steps, and the first and last stored frame of each replica supply the 20 DFT queries per round.

\FloatBarrier
The on-lattice aenet networks use Chebyshev descriptors and two five-unit tanh hidden layers, trained with the Broyden--Fletcher--Goldfarb--Shanno algorithm for up to 500 iterations with a 10\% test split. Descriptor settings accompany the Source Data. Final VASP free energies at the chosen electronic smearing define energies relative to each system's lowest-energy DFT configuration; only Nd$_2$Fe$_{14}$Co$_3$ has an ordered-parent reference under identical settings. The Monte Carlo temperatures guide configuration sampling rather than defining electronic smearing.

\subsection{Competing phases and matched references}

Energies above the convex hull use one eSEN checkpoint for candidates and competing phases, with the fixed-cell and cell-and-ion relaxation criteria in Table~\ref{SI-tabS:physical-settings}.

The primary catalogue contains 1,119 ordered three-dimensional Materials Project structures, without database-energy or stability filters. Each of 31 candidate systems admits all eligible subset-element phases and pooled candidates; database energies are unused. Adding three historical Materials Project-derived, 59 local OQMD and 50 local AFLOW structures leaves all 62 energies above the hull unchanged at reported precision. The local subsets are not complete database snapshots. One pure-fluorine reference is excluded; checks with alternative elemental-reference energies leave candidate energies above the hull unchanged. 

Calculation settings, PAW identifiers and energy conventions are supplied per candidate in the Source Data. New calculations preferentially use static energies of relaxed elemental references; previous results retain their matched relaxation references. The pooled calculations use either electronic free energies at the final relaxation step and chosen smearing or energies extrapolated to zero electronic smearing. Candidate and elemental-reference energies follow the same convention.

Matched-reference corrections use the Dudarev scheme \cite{dudarev1998lsdau}, with $U_{\rm eff}=(3.25,3.70,3.90,5.30,3.32,6.20)$~eV for V, Cr, Mn, Fe, Co and Ni. They are activated only when O or F and a listed $3d$ element are present. Elemental references use the same potential and $U_{\rm eff}$. Initial molecular O and F cells come from mp-1524462-r2SCAN (four O$_2$ molecules) and mp-1525632-r2SCAN (two F$_2$ molecules). Their full lattices, fractional coordinates and input settings are supplied in the Source Data. Both use PBE \cite{perdew1996pbe}, zero initial moments, 520~eV cutoff, 0.20~\AA$^{-1}$ reciprocal spacing, $10^{-6}$~eV electronic convergence and fixed cell shape and volume during ionic relaxation.

\FloatBarrier
\subsection{Rare-earth correction and calculation sensitivity}

Main-text Eqs.~\MainEq{eq:corrected-cell-moment}--\MainEq{eq:collinear-estimate} correct only PAW datasets \cite{blochl1994paw,kresse1999paw} omitting the specified $4f$ shell. Table~\ref{SI-tabS:open-core} lists the trivalent Hund-rule free-ion projections $g_JJ\mu_B$ with prescribed relative sublattice signs before the final absolute value \cite{miyake2018rareearth,liu2024ndfeco}. The corrected collinear saturation-magnetization estimate assumes saturated collinear sublattices and is distinct from a self-consistent magnetic ground state.

\begin{table}[!htbp]
\centering
\caption{Potential-specific open-core corrections used in the recorded calculations. Entries in the first column are the exact VASP PAW dataset identifiers retained for reproducibility. The total spin magnetic moment of the valence electrons in the entire cell and the corrections for the rare-earth $4f$ magnetic moments use the same sign convention, with the transition-metal sublattice moment taken as positive. Y$_{\rm sv}$ has no omitted $4f$ shell and receives no correction.}
\label{SI-tabS:open-core}
\small
\begin{tabular}{lrrrrrrr}
\toprule
PAW label & Assumed shell & $L$ & $S$ & $J$ & $g_JJ$ & Relative sign & $\Delta m/\mu_B$ \\
\midrule
Pr\_3 & $4f^2$ & 5 & 1 & 4 & $16/5$ & $+$ & $16/5$ \\
Nd\_3 & $4f^3$ & 6 & $3/2$ & $9/2$ & $36/11$ & $+$ & $36/11$ \\
Pm\_3 & $4f^4$ & 6 & 2 & 4 & $12/5$ & $+$ & $12/5$ \\
Sm\_3 & $4f^5$ & 5 & $5/2$ & $5/2$ & $5/7$ & $+$ & $5/7$ \\
Eu\_3 & $4f^6$ & 3 & 3 & 0 & 0 & N/A & 0 \\
Dy\_3 & $4f^9$ & 5 & $5/2$ & $15/2$ & 10 & $-$ & $-10$ \\
Ho\_3 & $4f^{10}$ & 6 & 2 & 8 & 10 & $-$ & $-10$ \\
Er\_3 & $4f^{11}$ & 6 & $3/2$ & $15/2$ & 9 & $-$ & $-9$ \\
Lu\_3 & $4f^{14}$ & 0 & 0 & 0 & 0 & N/A & 0 \\
Y\_sv & none & -- & -- & -- & 0 & N/A & 0 \\
\bottomrule
\end{tabular}
\end{table}
\FloatBarrier

For open-core Nd, the omitted $4f^3$ shell has $L=6$, $S=3/2$, $J=9/2$ and $g_J=8/11$, giving $g_JJ=36/11$ and approximately 3.27~$\mu_B$ per Nd.

For 21 available high-magnetization relax--static pairs, mean/maximum absolute differences are 1.24/6.49~meV/atom in formation energy and 9.81/28.11~mT in the corrected collinear saturation-magnetization estimate. Individual pairs accompany the Source Data.

\subsection{Compiled Nd-containing candidate collection}

Table~\ref{SI-tabS:pooled-dft} pools 62 reference-unseen Nd-containing structures with high corrected collinear saturation-magnetization estimates from the illustrative Nd-targeted run in main-text Fig.~\ref{fig:prospective}a,b and additional generation settings. This broader collection is not an independent evaluation set or a generation-yield estimate. The DFT formation energies and corrected collinear saturation-magnetization estimates remain distinct from eSEN energies above the convex hull, evaluated after separate cell-and-ion relaxation. Positive formation energies are retained, and the expanded competing-phase collections leave every displayed energy above the hull unchanged at the reported precision (Note~\ref{SI-sec:supp-note-4}).

\begingroup
\small
\setlength{\tabcolsep}{2pt}
\setlength{\LTcapwidth}{\textwidth}
\begin{longtable}{c@{\hspace{6pt}}cccccc}
\caption{Reference-unseen Nd-containing candidates from the illustrative Nd-targeted run in main-text Fig.~\ref{fig:prospective}a,b and additional generation settings. $E_{\rm form}^{\rm DFT}$ and $\mu_0M_s^{\rm col,est}$ are DFT-derived; $E_{\rm hull}^{\rm eSEN}$ is the energy above the convex hull from separate cell-and-ion relaxations and the frozen Materials Project competing-phase catalogue without database energies. Rows meet $\mu_0M_s^{\rm col,est}>1.5$~T, and positive matched-reference formation energies are retained. Energies are rounded for display; classification uses unrounded values. After removing repeated calculations, one representative is retained for each distinct combination of reduced formula, relaxed space-group number and computational-cell atom count $N$. Final DFT-relaxed structures and a row-matched index are available in the \href{https://github.com/shangwenchuang/GMMag/tree/main/results/dft_candidates_v1}{GMMag repository}.}\label{SI-tabS:pooled-dft}\\
\toprule
System & Formula & \shortstack{Space-group\\number} & $N$ & \shortstack{$\mu_0M_s^{\rm col,est}$\\(T)} & \shortstack{$E_{\mathrm{form}}^{\rm DFT}$\\(eV atom$^{-1}$)} & \shortstack{$E_{\mathrm{hull}}^{\rm eSEN}$\\(eV atom$^{-1}$)} \\
\midrule
\endfirsthead
\multicolumn{7}{c}{Table \thetable{} continued} \\
\toprule
System & Formula & \shortstack{Space-group\\number} & $N$ & \shortstack{$\mu_0M_s^{\rm col,est}$\\(T)} & \shortstack{$E_{\mathrm{form}}^{\rm DFT}$\\(eV atom$^{-1}$)} & \shortstack{$E_{\mathrm{hull}}^{\rm eSEN}$\\(eV atom$^{-1}$)} \\
\midrule
\endhead
\midrule\multicolumn{7}{r}{Continued on next page}\\
\endfoot
\bottomrule
\endlastfoot
Nd--Fe--Al & NdAlFe$_{9}$ & 31 & 22 & 1.715 & 0.158 & 0.233 \\
\addlinespace[1pt]
Nd--Fe--B & NdFe$_{10}$B & 3 & 12 & 1.865 & 0.059 & 0.135 \\
 & NdFe$_{7}$B & 65 & 18 & 1.759 & 0.313 & 0.429 \\
\addlinespace[1pt]
Nd--Fe--B--Eu & NdEuFe$_{14}$B & 136 & 68 & 1.730 & -0.076 & 0.048 \\
\addlinespace[1pt]
Nd--Fe--C--N & NdFe$_{15}$CN & 1 & 18 & 1.865 & 0.088 & 0.247 \\
\addlinespace[1pt]
Nd--Fe--Co & Nd$_{2}$Fe$_{14}$Co$_{3}$ & 166 & 57 & 1.990 & -0.001 & 0.032 \\
 & NdFe$_{7}$Co$_{4}$ & 127 & 24 & 1.970 & -0.006 & 0.048 \\
 & Nd$_{2}$Fe$_{13}$Co$_{4}$ & 160 & 57 & 1.941 & -0.001 & 0.043 \\
 & Nd$_{2}$Fe$_{12}$Co$_{5}$ & 12 & 57 & 1.898 & 0.003 & 0.057 \\
 & Nd$_{2}$Fe$_{12}$Co$_{5}$ & 166 & 57 & 1.898 & 0.003 & 0.057 \\
 & Nd$_{2}$Fe$_{11}$Co$_{6}$ & 166 & 57 & 1.884 & -0.024 & 0.040 \\
\addlinespace[1pt]
Nd--Fe--Co--Al & NdAl$_{3}$Fe$_{14}$Co & 8 & 57 & 1.530 & 0.016 & 0.166 \\
\addlinespace[1pt]
Nd--Fe--Co--Lu & NdLuFe$_{13}$Co$_{4}$ & 160 & 57 & 1.834 & -0.030 & 0.040 \\
\addlinespace[1pt]
Nd--Fe--Co--N & Nd$_{2}$Fe$_{14}$(CoN)$_{3}$ & 166 & 66 & 1.884 & -0.218 & 0.083 \\
\addlinespace[1pt]
Nd--Fe--Co--P & NdFe$_{14}$CoP$_{3}$ & 160 & 57 & 1.670 & -0.149 & 0.194 \\
 & NdFe$_{11}$Co$_{4}$P$_{3}$ & 160 & 57 & 1.557 & -0.126 & 0.241 \\
\addlinespace[1pt]
Nd--Fe--Co--Pm & PmNdFe$_{14}$Co$_{3}$ & 194 & 38 & 1.949 & -0.007 & 0.024 \\
\addlinespace[1pt]
Nd--Fe--Co--Pr & PrNdFe$_{14}$Co$_{3}$ & 194 & 38 & 1.973 & 0.005 & 0.037 \\
 & PrNdFe$_{13}$Co$_{4}$ & 160 & 57 & 1.933 & 0.005 & 0.048 \\
 & PrNdFe$_{12}$Co$_{5}$ & 40 & 38 & 1.889 & 0.010 & 0.064 \\
\addlinespace[1pt]
Nd--Fe--Co--Rh & NdFe$_{13}$Co$_{3}$Rh$_{2}$ & 187 & 38 & 1.846 & 0.116 & 0.248 \\
\addlinespace[1pt]
Nd--Fe--Co--Si & NdFe$_{14}$CoSi$_{3}$ & 160 & 57 & 1.624 & -0.109 & 0.148 \\
\addlinespace[1pt]
Nd--Fe--Co--Sm & NdSmFe$_{16}$Co & 160 & 57 & 1.903 & 0.032 & 0.041 \\
 & NdSmFe$_{14}$Co$_{3}$ & 194 & 38 & 1.877 & -0.011 & 0.024 \\
 & NdSmFe$_{15}$Co$_{2}$ & 194 & 38 & 1.875 & 0.033 & 0.056 \\
 & NdSmFe$_{14}$Co$_{3}$ & 160 & 57 & 1.854 & -0.011 & 0.024 \\
 & NdSmFe$_{13}$Co$_{4}$ & 160 & 57 & 1.836 & -0.011 & 0.036 \\
 & NdSmFe$_{12}$Co$_{5}$ & 160 & 57 & 1.794 & -0.007 & 0.050 \\
 & NdSmFe$_{11}$Co$_{6}$ & 194 & 38 & 1.778 & -0.040 & 0.022 \\
 & NdSmFe$_{10}$Co$_{7}$ & 160 & 57 & 1.730 & -0.038 & 0.034 \\
\addlinespace[1pt]
Nd--Fe--Co--Sm--V & NdSmVFe$_{15}$Co & 160 & 57 & 1.614 & 0.002 & 0.042 \\
\addlinespace[1pt]
Nd--Fe--Co--Y & NdYFe$_{16}$Co & 160 & 57 & 1.878 & 0.017 & 0.041 \\
 & NdYFe$_{14}$Co$_{3}$ & 194 & 38 & 1.861 & -0.026 & 0.022 \\
 & NdYFe$_{14}$Co$_{3}$ & 160 & 57 & 1.838 & -0.024 & 0.023 \\
 & NdYFe$_{13}$Co$_{4}$ & 160 & 57 & 1.804 & -0.025 & 0.032 \\
 & NdYFe$_{11}$Co$_{6}$ & 194 & 38 & 1.761 & -0.056 & 0.014 \\
\addlinespace[1pt]
Nd--Fe--Co--Zr & NdZrFe$_{14}$Co$_{3}$ & 160 & 57 & 1.870 & -0.022 & 0.055 \\
 & NdZrFe$_{13}$Co$_{4}$ & 160 & 57 & 1.825 & -0.023 & 0.065 \\
 & NdZrFe$_{16}$Co & 160 & 57 & 1.814 & 0.011 & 0.065 \\
 & NdZrFe$_{15}$Co$_{2}$ & 160 & 57 & 1.793 & 0.014 & 0.081 \\
 & NdZrFe$_{11}$Co$_{6}$ & 194 & 38 & 1.767 & -0.060 & 0.038 \\
 & NdZrFe$_{12}$Co$_{5}$ & 160 & 57 & 1.763 & -0.026 & 0.071 \\
\addlinespace[1pt]
Nd--Fe--Cr--Zr & NdZrCrFe$_{16}$ & 160 & 57 & 1.532 & 0.005 & 0.044 \\
\addlinespace[1pt]
Nd--Fe--Ga--Sm & NdSmGaFe$_{16}$ & 160 & 57 & 1.709 & 0.006 & 0.065 \\
\addlinespace[1pt]
Nd--Fe--Lu & NdLuFe$_{17}$ & 160 & 57 & 1.912 & 0.013 & 0.032 \\
 & NdLuFe$_{17}$ & 194 & 38 & 1.891 & 0.016 & 0.033 \\
\addlinespace[1pt]
Nd--Fe--N & NdFe$_{8}$N$_{3}$ & 1 & 24 & 1.532 & -0.051 & 0.251 \\
\addlinespace[1pt]
Nd--Fe--O & NdFe$_{8}$O & 1 & 20 & 1.814 & -0.269 & 1.025 \\
\addlinespace[1pt]
Nd--Fe--P & NdFe$_{13}$P$_{2}$ & 1 & 16 & 1.823 & -0.018 & 0.276 \\
 & NdFe$_{9}$P & 6 & 22 & 1.812 & 0.028 & 0.309 \\
 & Nd(Fe$_{5}$P)$_{3}$ & 160 & 57 & 1.695 & -0.155 & 0.173 \\
 & Nd(Fe$_{6}$P)$_{2}$ & 6 & 15 & 1.629 & 0.051 & 0.219 \\
\addlinespace[1pt]
Nd--Fe--Rh & Nd$_{2}$Fe$_{15}$Rh$_{2}$ & 166 & 57 & 1.908 & 0.016 & 0.154 \\
\addlinespace[1pt]
Nd--Fe--S & NdFe$_{11}$S$_{3}$ & 173 & 30 & 1.563 & -0.036 & 0.416 \\
\addlinespace[1pt]
Nd--Fe--Si & Nd(Fe$_{5}$Si)$_{3}$ & 160 & 57 & 1.686 & -0.107 & 0.127 \\
\addlinespace[1pt]
Nd--Fe--Sm--V & NdSmVFe$_{16}$ & 160 & 57 & 1.654 & 0.000 & 0.025 \\
\addlinespace[1pt]
Nd--Fe--Ti & NdTiFe$_{17}$ & 164 & 38 & 1.838 & 0.057 & 0.099 \\
\addlinespace[1pt]
Nd--Fe--Zr & NdZrFe$_{17}$ & 194 & 38 & 1.884 & 0.007 & 0.045 \\
 & NdZrFe$_{17}$ & 160 & 57 & 1.857 & 0.013 & 0.054 \\
 & NdZrFe$_{10}$ & 47 & 12 & 1.545 & -0.023 & 0.050 \\
\addlinespace[1pt]
Nd--Cl--Cr & NdCr$_{2}$Cl & 59 & 8 & 1.521 & 0.215 & 0.967 \\
\addlinespace[1pt]
Nd--Cr--F & NdCr$_{2}$F & 8 & 8 & 1.625 & -1.026 & 2.018 \\
\end{longtable}
\relax
\endgroup
\FloatBarrier
\relax
\FloatBarrier
\relax
\putbib[references]
\end{bibunit}

\end{document}